\documentclass[aps,prc,reprint,superscriptaddress]{revtex4-2}

\usepackage{graphicx}            % Include graphics
\usepackage{amsmath,amssymb}     % Mathematical tools for equations and symbols
\usepackage{float}               % Enhanced floating environments (figures, tables)
\usepackage{dcolumn}
\usepackage{longtable}
\begin{document}

%\preprint{}

    \title{Study of $\boldsymbol\beta$ Decay Shape Factors in First-Forbidden Transitions with $\boldsymbol{\Delta I^\pi = 0^-}$ for Reactor Antineutrino Spectra Predictions}

    \author{G. A. Alcal\'a}
    \email[]{galcala@ific.uv.es}
    \affiliation{Instituto de F\'isica Corpuscular, CSIC-Universitat de Val\`encia, E-46071 Val\`encia, Spain}
    
    \author{A. Algora}
    \email[]{algora@ific.uv.es}
    \affiliation{Instituto de F\'isica Corpuscular, CSIC-Universitat de Val\`encia, E-46071 Val\`encia, Spain}
    \affiliation{HUN-REN Institute for Nuclear Research (ATOMKI), H-4001 Debrecen, Hungary}
    
    \author{M. Estienne}
    \affiliation{SUBATECH, CNRS/IN2P3, IMT Atlantique, Nantes Université, F-44307 Nantes, France}
    
    \author{M. Fallot}
    \affiliation{SUBATECH, CNRS/IN2P3, IMT Atlantique, Nantes Université, F-44307 Nantes, France}
    
    \author{V. Guadilla}
    \affiliation{SUBATECH, CNRS/IN2P3, IMT Atlantique, Nantes Université, F-44307 Nantes, France}
    \affiliation{Faculty of Physics, University of Warsaw, 02-093 Warsaw, Poland.}
      
    \author{A. Beloeuvre}
    \affiliation{SUBATECH, CNRS/IN2P3, IMT Atlantique, Nantes Université, F-44307 Nantes, France}
    
    \author{W. Gelletly}
    \affiliation{Department of Physics, University of Surrey, GU27XH Guildford, United Kingdom}
     
    \author{R. Kean}
    \affiliation{SUBATECH, CNRS/IN2P3, IMT Atlantique, Nantes Université, F-44307 Nantes, France}
    
    \author{A. Porta}
    \affiliation{SUBATECH, CNRS/IN2P3, IMT Atlantique, Nantes Université, F-44307 Nantes, France}
    
    \author{S. Bouvier}
    \affiliation{SUBATECH, CNRS/IN2P3, IMT Atlantique, Nantes Université, F-44307 Nantes, France}
    
    \author{J.-S. Stutzmann}
    \affiliation{SUBATECH, CNRS/IN2P3, IMT Atlantique, Nantes Université, F-44307 Nantes, France}
    
    \author{E. Bonnet}
    \affiliation{SUBATECH, CNRS/IN2P3, IMT Atlantique, Nantes Université, F-44307 Nantes, France}
    
    \author{T. Eronen}
    \affiliation{Department of Physics, University of Jyväskylä, P.O. Box 35, FI-40014 Jyväskylä, Finland}
    
    \author{D. Etasse}
    \affiliation{LPC Caen, ENSICAEN, Université de Caen, CNRS/IN2P3, F-Caen, France}
    
    \author{J. Agramunt}
    \affiliation{Instituto de F\'isica Corpuscular, CSIC-Universitat de Val\`encia, E-46071 Val\`encia, Spain}
    
    \author{J. L. Tain}
    \affiliation{Instituto de F\'isica Corpuscular, CSIC-Universitat de Val\`encia, E-46071 Val\`encia, Spain}
    
    \author{H. Garcia Cabrera}
    \affiliation{Instituto de F\'isica Corpuscular, CSIC-Universitat de Val\`encia, E-46071 Val\`encia, Spain}
    \affiliation{Universidad Europea de Madrid, Department of Biosciences, Faculty of Biomedical and Health Sciences, E-28670 Madrid, Spain}
    
    \author{L. Giot}
    \affiliation{SUBATECH, CNRS/IN2P3, IMT Atlantique, Nantes Université, F-44307 Nantes, France}
    
    \author{A. Laureau}
    \affiliation{SUBATECH, CNRS/IN2P3, IMT Atlantique, Nantes Université, F-44307 Nantes, France}
    
    \author{J. A. Victoria}
    \affiliation{Instituto de F\'isica Corpuscular, CSIC-Universitat de Val\`encia, E-46071 Val\`encia, Spain}
    
    \author{Y. Molla}
    \affiliation{SUBATECH, CNRS/IN2P3, IMT Atlantique, Nantes Université, F-44307 Nantes, France}
    
    \author{A. Jaries}
    \affiliation{Department of Physics, University of Jyväskylä, P.O. Box 35, FI-40014 Jyväskylä, Finland}
    
    \author{L. Al Ayoubi}
    \affiliation{Department of Physics, University of Jyväskylä, P.O. Box 35, FI-40014 Jyväskylä, Finland}
    
    \author{O. Beliuskina}
    \affiliation{Department of Physics, University of Jyväskylä, P.O. Box 35, FI-40014 Jyväskylä, Finland}
    
    \author{W. Gins}
    \affiliation{Department of Physics, University of Jyväskylä, P.O. Box 35, FI-40014 Jyväskylä, Finland}
    
    \author{M. Hukkanen}
    \affiliation{Department of Physics, University of Jyväskylä, P.O. Box 35, FI-40014 Jyväskylä, Finland}
    
    \author{A. Illana}
    \affiliation{Department of Physics, University of Jyväskylä, P.O. Box 35, FI-40014 Jyväskylä, Finland}
    \affiliation{Grupo de Física Nuclear, EMFTEL, and IPARCOS, Universidad Complutense de Madrid, CEI Moncloa, E-28040 Madrid, Spain.}
    
    \author{A. Kankainen}
    \affiliation{Department of Physics, University of Jyväskylä, P.O. Box 35, FI-40014 Jyväskylä, Finland}
    
    \author{S. Kujanpää}
    \affiliation{Department of Physics, University of Jyväskylä, P.O. Box 35, FI-40014 Jyväskylä, Finland}
    
    \author{I. Moore}
    \affiliation{Department of Physics, University of Jyväskylä, P.O. Box 35, FI-40014 Jyväskylä, Finland}
    
    \author{I. Pohjalainen}
    \affiliation{Department of Physics, University of Jyväskylä, P.O. Box 35, FI-40014 Jyväskylä, Finland}
    
    \author{D. Pitman}
    \affiliation{Department of Physics, University of Jyväskylä, P.O. Box 35, FI-40014 Jyväskylä, Finland}
    \affiliation{Department of Physics and Astronomy, University of Manchester, Manchester M13 9PL, United Kingdom}
    
    \author{A. Raggio}
    \affiliation{Department of Physics, University of Jyväskylä, P.O. Box 35, FI-40014 Jyväskylä, Finland}
    
    \author{M. Reponen}
    \affiliation{Department of Physics, University of Jyväskylä, P.O. Box 35, FI-40014 Jyväskylä, Finland}

    \author{J. Romero}
    \affiliation{Department of Physics, University of Jyväskylä, P.O. Box 35, FI-40014 Jyväskylä, Finland}
    \affiliation{Oliver Lodge Laboratory, University of Liverpool, Liverpool, L69 7ZE, United Kingdom}
    
    \author{J. Ruotsalainen}
    \affiliation{Department of Physics, University of Jyväskylä, P.O. Box 35, FI-40014 Jyväskylä, Finland}
    
    \author{M. Stryjczyk}
    \affiliation{Department of Physics, University of Jyväskylä, P.O. Box 35, FI-40014 Jyväskylä, Finland}
    \affiliation{Institut Laue-Langevin, 71 Avenue des Martyrs, F-38042 Grenoble, France}

    \author{A. Tolosa}
    \affiliation{Department of Physics, University of Jyväskylä, P.O. Box 35, FI-40014 Jyväskylä, Finland}
    \affiliation{European Organization for Nuclear Research (CERN), Geneva, Switzerland}
    
    \author{V. Virtanen}
    \affiliation{Department of Physics, University of Jyväskylä, P.O. Box 35, FI-40014 Jyväskylä, Finland}

    \date{\today}

    \begin{abstract}
 %       Electron $\Delta$E–E detectors were developed and employed by the \textit{e-Shape} collaboration to measure the shapes of relevant $\beta$ decay spectra to determine appropriate $\beta$ feedings and shape models for reactor antineutrino spectrum predictions. 
 The electron spectra of the $\beta$ decays of $^{92}$Rb and $^{142}$Cs, key contributors to the reactor antineutrino spectrum, were measured at the IGISOL facility using radioactive beams of high isotopic purity. The shapes of the measured $\beta$ spectra were compared with various $\beta$ shape models, including first-forbidden correction factors for $\Delta I^\pi = 0^-$ ground-state to ground-state transitions. Comparisons with previous experimental results are also provided. The shapes of the newly measured $\beta$ spectra are well reproduced employing feedings extracted from total absorption gamma spectroscopy measurements. %This article extends the results previously presented in \textit{Physical Review Letters.
    \end{abstract}

%\keywords{}

    \maketitle

    \section{Introduction}
    
        Nuclear reactors are the most powerful human-made sources of electron antineutrinos ($\bar{\nu}_e$) on Earth. These facilities are crucial for studying general properties of neutrinos, such as flavor oscillation phenomena and their mass hierarchy \cite{MENTION}. Reactor neutrino research is also relevant for security matters, since neutrino detectors can be used, in principle, for nonproliferation control of nuclear facilities, power monitoring, as well as for supervision of spent fissile fuel \cite{Bernstein}. From a more fundamental perspective, studying the shape of the reactor antineutrino spectrum through the contributing beta ($\beta$) spectra is important for weak interaction research, including a search for physics beyond the Standard Model \cite{Rozpedzik_2023,DeKeukeleere_2024}.

        Accurate measurements and reliable methods for calculating reactor antineutrino spectra predictions are essential tools for the research topics mentioned above. Two basic approaches are used to obtain these predictions: the conversion method and the summation method \cite{SCHRECKENBACH1,KING_1958}. Both require detailed nuclear data information, including the shapes of $\beta$ spectra. In this article, we focus on the precise determination of $\beta$ spectra that significantly contribute to the reactor antineutrino spectra to improve such predictions. The present work gives a more detailed description of our recently published Letter \cite{ALCALA_PRL}.

        In the conversion method, the reactor antineutrino spectrum is obtained by reproducing the experimentally obtained aggregate $\beta$ spectra of every of the individual fissile isotopes in the reactor fuel ($^{239}$Pu, $^{241}$Pu, $^{235}$U, and $^{238}$U). The total antineutrino spectrum of the reactor is then calculated as the weighted sum of the deduced aggregate antineutrino spectra of these fissile isotopes, derived from the reproductions of their experimental aggregate $\beta$ spectra. The weights correspond to the respective fractions of the fissile isotopes in the reactor fuel composition \cite{SCHRECKENBACH1,SCHRECKENBACH2,HUBER,MUELLER}. The reproduction of an aggregate $\beta$ spectrum of a particular fissile isotope is performed using a reduced number of average virtual $\beta$ spectra for effective atomic number values (average $Z$). The shapes of the virtual $\beta$ spectra are approximated to those of allowed transitions, and their weights are adjusted to the experimental aggregate spectrum. The corresponding antineutrino spectra related to the adjusted virtual $\beta$ spectra are then calculated based on energy conservation (between the electron–antineutrino pair) and subsequently summed.
        
        The main caveat of the conversion method is in the use of average virtual $\beta$ spectra. The conversion from the virtual spectra to antineutrino is an approximation, since the exact conversions should be carried out for each of the individual real $\beta$ transitions associated with every individual $\beta$ branch of the decay of the isotopes produced in a reactor. Rigorously speaking, the energy conservation relation holds true only for these individual transitions. Moreover, any assumptions about the shapes of the virtual $\beta$ spectra also affect the calculated antineutrino spectrum.

        An alternative option is the summation method \cite{KING_1958,AVIGNONE} (also refereed as \textit{ab initio} approach in \cite{MUELLER}), in which an aggregate $\beta$ spectrum is constructed by summing all contributing individual $\beta$ spectra, following the relation
        \begin{equation}
            S_k = \sum_{i,j} A_{k,i} \, f_{i,j} \, s_{i,j} \; ,
            \label{eq_summation}
        \end{equation}
        \noindent where $S_k$ is the aggregate $\beta$ spectrum of the $k$ fissile isotope, $A_{k,i}$ is the activity of the $i$ fission product or subsequent parent nucleus produced in the reactor, and $f_{i,j}$ is the $\beta$ feeding (or $\beta$ transition probability) of the $s_{i,j}$ $\beta$ spectrum for a particular $\beta$ transition to the $j$ energy level of the respective daughter nucleus. The activity of the parent nuclei is associated with the yield of the fission fragments of the reactor fuel.

        The conversion method has been commonly used to predict the measured antineutrino spectra from reactors in many important experiments through the years \cite{MENTION,RENO,DAYABAY}. The aggregate spectra measured by \cite{SCHRECKENBACH1,SCHRECKENBACH2,VONFEILITZSCH_1982,SCHRECKENBACH3} for $^{239}$Pu, $^{241}$Pu, and $^{235}$U have been used as a reference for these calculations, given the absence of more precise measurements. In what follows, the commented spectra will be referred to as the Schreckenbach data. The corresponding $^{238}$U aggregate spectrum is also required but was not measured by Schreckenbach and his collaborators. Accordingly, it was obtained previously from summation calculations \cite{MUELLER}. More recently, the aggregate $\beta$ spectrum of the fission of $^{238}$U has been measured \cite{HAAG}, and thus the associated antineutrino spectrum is also available to be used with this method.

        Given the need for reliable calculations of the antineutrino spectra to reproduce the experimental data collected at several reactors worldwide \cite{DCHOOZ2,RENO,DAYABAY}, Huber \cite{HUBER} and Mueller \textit{et al.} \cite{MUELLER} revisited the conversion calculations. Their revisions employed updated nuclear databases and implemented several improvements. Huber \cite{HUBER} used a pure conversion method with the best nuclear data available in ENSDF \cite{ENSDF} at that time, and utilized a $\beta$ shape model with allowed shape correction factors. Mueller \textit{et al.} \cite{MUELLER} also used the nuclear data available at ENSDF \cite{ENSDF} and employed a hybrid procedure to calculate the antineutrino spectrum. In their approach, initially a summation method was used to generate an aggregate $\beta$ spectrum, which was incomplete since it only included well-known individual $\beta$ spectra. Then, the conversion method was applied to account for the differences between this incomplete $\beta$ spectrum and the experimental one \cite{SCHRECKENBACH1,SCHRECKENBACH2,VONFEILITZSCH_1982,SCHRECKENBACH3}. In both approaches, the generated spectra are constrained by the Schreckenbach data. Revisions of models related to antineutrino spectra calculations, together with the works of Huber and Mueller \textit{et al.}, revealed a difference between the experimental and predicted antineutrino flux in reactors at short baselines \cite{MENTION} (defined as measurements at distances less than 100~m). This discrepancy has been referred to as the ``Reactor Antineutrino Anomaly'' (RAA).

        The RAA is a 6\% deficit in the measured antineutrino flux compared to predictions for several reactor experiments worldwide \cite{MENTION}. Possible sources of this anomaly have been considered and explored by the physics community. Several causes have been proposed, such as unidentified problems in the detection of antineutrinos due to the design of the detectors, or inadequacies in data management, including unaccounted systematic errors. Other possible sources include an erroneous characterization of the reactor fuel composition, or the hypothetical existence of a fourth neutrino flavor possessing an even weaker interaction with matter, referred to as sterile neutrinos. Another possibility involves the effects of incomplete nuclear databases and nuclear models used as inputs in the calculations.

        The works of Fallot \textit{et al.} \cite{FALLOT} and Estienne \textit{et al.} \cite{ESTIENNE}, based on the summation method, provide solid evidence for the role of incomplete nuclear data. The authors employed a pure summation method to calculate reactor antineutrino spectra, critically selecting the $\beta$ decay data of the most relevant fission fragments. The $\beta$ decay data employed were selected in priority order, giving preference to data that avoid the Pandemonium effect \cite{HARDY}. Therefore, for some of these relevant decays, $\beta$ feedings have been determined using the Total Absorption Gamma Spectroscopy (TAGS) technique \cite{ALGORA1,TAGSBOOK}.
        
        To understand the use of TAGS $\beta$ feedings to avoid the Pandemonium effect, it must first be considered that it is quite common to find sets of $\beta$ feedings obtained with High-Purity Germanium (HPGe) detectors (High-Resolution Gamma Spectroscopy) in the literature. HPGe detectors are characterized by having low efficiencies for registering weak and/or high-energy gamma-rays ($\gamma$). Since the feedings are determined from the intensity balance populating and de-exciting a nuclear level, feedings can be erroneously assigned due to the possible non-detection of some $\gamma$-rays, producing an overestimation of the feedings at low excitation energies in the daughter nucleus (a systematic error). This systematic error in the feedings is what is called the Pandemonium effect \cite{HARDY}. This effect has been well studied with the TAGS technique, which relies on the use of high-efficiency calorimeter detectors \cite{ALGORA1}.

        Estienne \textit{et al.} \cite{ESTIENNE} reported a systematic reduction in the reactor flux anomaly to 1.9\% with the progressive inclusion of $\beta$ feedings measured by TAGS. Therefore, predictions calculated using the summation method, including TAGS data, demonstrate that the RAA is mostly associated with the limitations of the conversion method. It is worth highlighting that the summation methodology has successfully reduced the RAA without needing complete knowledge of the decay properties of all unstable nuclei produced in a reactor. This is because only a relatively small number of nuclei contribute significantly to the total reactor antineutrino spectrum. Their contributions are determined by the fission yields of the respective fission fragments and the production of their subsequent daughter nuclei \cite{ESTIENNE,ALGORA1,ZAKARI}. Therefore, the inclusion of more accurate $\beta$ feedings for these particular nuclei has a significant impact on the summation calculations.
        
        It has also been proposed that the RAA may be further explained through the use of accurate shape models for $\beta$ spectra that include appropriate first-forbidden shape correction factors \cite{HAYES2}, since approximately 25\% of $\beta$ transitions in reactor antineutrino spectra are forbidden. Recent works by Kopeikin \textit{et al.} \cite{KOPEIKIN_PyhsAN,KOPEIKIN} demonstrate that the normalization of the $^{235}$U aggregate $\beta$ spectrum measured at the ILL reactor \cite{SCHRECKENBACH1,SCHRECKENBACH2} was overestimated by about 5\%. Correcting this normalization reduces the predicted antineutrino spectrum to a level where the RAA becomes statistically insignificant. This conclusion is consistent with and reinforces the results of Estienne \textit{et al.} \cite{ESTIENNE}, whose determination of reactor antineutrino spectra does not rely on measurements of aggregate $\beta$ spectra. It is worth mentioning that Fallot \textit{et al.} \cite{FALLOT_PRL2012} and Estienne \textit{et al.} \cite{ESTIENNE} identified inconsistencies in the normalization of the predictions calculated by the summation method and Huber \cite{HUBER} for the $^{235}$U aggregate spectrum, which can be explored with Kopeikin \textit{et al.} \cite{KOPEIKIN_PyhsAN,KOPEIKIN} data. 

        Another discrepancy found in reactor spectra, which is yet to be understood, is the shape anomaly. While the RAA is considered a normalization problem of the experimental data, the shape anomaly corresponds to a deformation of the energy spectra, characterized by a ``shoulder'' or ``bump'' in the experimental data around the 6 MeV kinetic energy region of reactor antineutrino spectra when compared with theoretical predictions \cite{MENTION}. Hayes \textit{et al.} \cite{HAYES1} discussed several possible causes for this deformation, ranging from contributions to the spectra of neutron-induced reactions of non-fuel materials inside reactors, to the effects of different neutron fluxes, or to possible errors in the measurement of aggregate spectra. Another proposed cause is related to the forbidden nature of the fission fragment $\beta$ spectra, which is unaccounted for in the predictions. This last hypothesis requires further investigation, which is the goal of the present work.

        Hayen \textit{et al.} \cite{HAYEN_1F} considered that the inclusion of proper first-forbidden correction factors in the calculated $\beta$ transitions that dominate the energy region of the ``bump'' can be a relevant step to solve the discrepancy. Their claim is based on the fact that the approximations used for non-unique forbidden transitions in previous calculations of antineutrino spectra were not appropriate, as many were carried out using allowed $\beta$ shapes. Forbidden transitions have a significant influence on the total antineutrino spectrum, since the energy region of the ``bump'' is dominated by these types of transitions rather than allowed ones. Therefore, Hayen \textit{et al.} \cite{HAYEN_1F} calculated first-forbidden shape correction factors for non-unique $\beta$ transitions with dominant contributions to the reactor spectrum in the ``bump'' region. These corrections were based on the formalism of Behrens and B\"uhring \cite{BEHRENS_1971,BEHRENS_BOOK}.

        The unknown origin of the ``bump'' demands accurate measurements of individual $\beta$ spectra to clarify the causes of the reactor antineutrino spectrum shape deformation \cite{HAYES1,HAYEN_1F}. To carry out these measurements, the \textit{e-Shape} collaboration developed and built new electron telescope ($\Delta$E–E) detectors \cite{GUADILLA} to measure the shapes of $\beta$ decay spectra of isotopes that contribute most to reactor antineutrino spectra. To describe these measurements theoretically, it will be necessary to select the most appropriate sets of $\beta$ feedings (either from TAGS or high-resolution spectroscopy) and use accurate allowed \cite{HAYEN_Allowed,HUBER} and first-forbidden \cite{HAYEN_1F} shape correction factors. Correct feedings and shape models are essential for generating reliable predictions of reactor antineutrino spectra with the summation method. Consequently, in this article, we present the measurement and analysis of the $^{92}$Rb and $^{142}$Cs $\beta$ decay spectra, as these two are respectively the first and third largest contributors to the aggregate spectrum of the fission of the $^{235}$U \cite{SONZOGNI1}.
        
        This work first introduces the challenges and possible solutions related to reactor antineutrino spectra. Next, the allowed and non-unique first-forbidden shape correction factors to $\beta$ spectra, which are expected to improve predictions of reactor energy spectra, are presented. A full description of the design of the electron $\Delta$E–E detectors developed for measuring relevant $\beta$ spectra is then given, together with a detailed account of the I233 experimental campaign carried out at the IGISOL facility (University of Jyv\"askyl\"a, Finland) for measuring these spectra. Finally, the preparation of the data and the analysis tools for the $^{92}$Rb and $^{142}$Cs $\beta$ spectra are described, along with comparisons to previous experimental results \cite{TENGBLAD,RUDSTAM_1990}.

    \section{$\boldsymbol\beta$ spectra shape corrections}

        Reactor fission fragments are neutron-rich nuclei that decay via the emission of electron–antineutrino pairs. Depending on the changes in nuclear spin and parity, the decay can, in general, be classified as allowed or forbidden, with several subclassifications that depend on the angular momentum carried by the electron–antineutrino pair. In particular, the shape of the energy spectrum of the electron (and antineutrino) is sensitive to changes in nuclear properties between the initial and final nuclear states of the involved nuclei, the charge state of the system, and even the atomic and molecular conditions when the electron is emitted. Therefore, the shape of the $\beta$ decay energy spectrum for a particular $\beta$ transition, from the initial nuclear energy level of the parent nucleus to the final energy level of the daughter nucleus, is written in natural units as
        \begin{equation}
            \label{eq_betaspec}
            S_\beta(W) \propto \eta W (W_0-W)^2 F(W,Z_d,A) C(Z_d,W) K(Z_d,W) \;,
        \end{equation}
        \noindent where $\eta$ is the normalized momentum of the electron ($\eta=p_e/m_ec$), $W$ is the normalized total energy of the electron ($W=E_e/m_ec^2$), $W_0$ is the maximum total energy available for the particles in the transition ($W_0=Q_e/m_ec^2$, being $Q_e$ the $Q$ value of the reaction), $Z_d$ is the atomic number of the daughter nucleus, and $A$ is the mass number of both the parent and daughter nuclei. The term $\eta W (W_0-W)^2$ is known as the statistical factor, derived from Fermi’s theory of $\beta$ decay, and represents the basic shape of a $\beta$ decay spectrum. $F(W,Z_d,A)$ is the Fermi function \cite{EVANS}, which, to first order, accounts for the Coulomb effect of the nuclear charge of the daughter nucleus on the emitted electrons. $C(Z_d,W)$ is the nuclear shape factor. It is sensitive to changes in the nuclear properties as a consequence of the $\beta$ transition, and encodes information about the nuclear structure of the decaying system. For allowed transitions, our work uses the common approximation $C(Z_d,W)=1$, which assumes that the nuclear matrix elements of these transitions are independent of the energy of the emitted lepton pair. For non-unique first-forbidden transitions, we employ the calculations of Hayen \textit{et al.} \cite{HAYEN_1F}. $K(Z_d,W)$ accounts for higher-order corrections associated with additional nuclear, atomic, and molecular effects not included in the previous shape factors. In this work, the sets of allowed correction factors proposed by Huber \cite{HUBER} and Hayen \textit{et al.} \cite{HAYEN_Allowed} were tested in the comparisons between measured and predicted $\beta$ spectra.

        The allowed shape corrections of Huber \cite{HUBER} are a set of shape factors selected for their relevance in the energy region where the anomalies of the reactor spectrum are most pronounced. As a standard, the statistical factor and the Fermi function must always be included in the modeling of any $\beta$ spectrum, as these are ubiquitous to all $\beta$ transitions and are the most dominant shape factors. In addition to the standard shape, Huber \cite{HUBER} includes several higher-order correction factors, starting with two corrections for the finite size of the nucleus. The first, called the ``finite size of the nucleus'' correction, accounts for the fact that nuclei are not point-like objects. The second, named the ``weak interaction finite size of the nucleus'' correction, is similar to the former, but it is formulated from the weak-interaction perspective and is valid only for Gamow-Teller transitions. An ``atomic screening'' correction was included to account for the effect of the atomic electrons on the electrostatic field of the nuclear charge. A ``radiative'' correction was implemented to account for higher-order electromagnetic interactions between the $\beta$ electrons and the nuclear charge that are not included in the Fermi function. Finally, the ``weak magnetism'' correction is applied only to Gamow-Teller transitions to account for the effect of induced currents that do not appear in the initial Hamiltonian of the V-A theory \cite{GELLMANN}.

        Hayen \textit{et al.} \cite{HAYEN_Allowed} propose a more general set of shape correction factors that can be used to model $\beta$ spectra over a wider energy range. They present 19 possible shape correction factors, grouped in subsets that account for different degrees of relevance in a $\beta$ spectrum. In our work, the most relevant factors of Hayen \textit{et al.} \cite{HAYEN_Allowed} were selected according to their magnitudes and the expected characteristics of the $\beta$ decays that contribute to the reactor spectrum. The first subset of Hayen \textit{et al.} \cite{HAYEN_Allowed} is formed by the statistical factor and the Fermi function. This subset must be included in our work since its corrections are listed as the most relevant factors, with contributions to a $\beta$ spectrum above the percentage level. The second subset of corrections is reported to have effects of order $10^{-1}$ to $10^{-2}$ relative to the first subset. From this subset, the ``finite size of the nucleus'' and the ``radiative'' corrections were included in our predictions. It must be noted that the latter contains higher-order terms than the one given by Huber \cite{HUBER}. From this same subset, the ``atomic exchange'' and the ``atomic mismatch'' corrections were also added. The former atomic corrections account for the simultaneous emission of atomic electrons during the decay, which are indistinguishable from the $\beta$ electrons, and the latter for the loss of kinetic energy of $\beta$ electrons due to excitations of atomic electrons. These atomic effects are related to overlaps between the atomic and nuclear wave functions. The third subset of corrections reports magnitudes of order $10^{-3}$ to $10^{-4}$. From this subset, the ``atomic screening'' correction was included (noting that its mathematical expression differs from the one presented by Huber \cite{HUBER}). From the same subset, the ``recoil of the nucleus'' and the ``distorted Coulomb potential due to recoil'' corrections were also added. The former is a kinematic correction that accounts for the energy loss of $\beta$ electrons due to the recoil motion of the daughter nucleus, while the latter corrects the nuclear electrostatic potential to account for the displacement of the nucleus due to recoil.

	\section{$\boldsymbol\Delta$E–E detectors}

		Measurements of $\beta$-decay spectra were performed with electron telescope ($\Delta E$–$E$) detectors developed by the \textit{e-Shape} collaboration. The setup is composed of two detectors placed at 70$^\circ$ relative to the axis passing through their geometric centers (i.e., 35$^\circ$ with respect to the direction of the radioactive beam). These detectors can be used either under vacuum or at atmospheric pressure. The design is conceptually similar to detectors used in earlier works for measuring $\beta$ spectra \cite{TENGBLAD,HOROWITZ1994}.
        
        The \textit{e-Shape} detectors are electron spectrometers designed for $\gamma$-ray rejection in coincidence mode \cite{GUADILLA}. The $\Delta$E part is a non-segmented thin layer of high-purity silicon (Si detector) manufactured by Micron (MSX25 model). The E part is a thick plastic scintillator. The Si detector was cut into a square shape of 53.78$\times$53.78 mm$^2$, with an active area of 50$\times$50 mm$^2$, a thickness of 542 $\mu$m, and a superficial dead layer of the order of 1 $\mu$m. The plastic detector is made of EJ-200 scintillation material produced by Scionix. It is shaped as a 7.5 cm high truncated pyramid (tapered shape), with a rounded base for coupling to a photomultiplier (PMT) and a top face cut to match the active area of the Si detector. The thickness of the plastic guarantees the full kinetic energy deposition of electrons up to 10 MeV.  
        
\begin{figure}[h]
		    \centering
		    \includegraphics[width=0.475\textwidth]{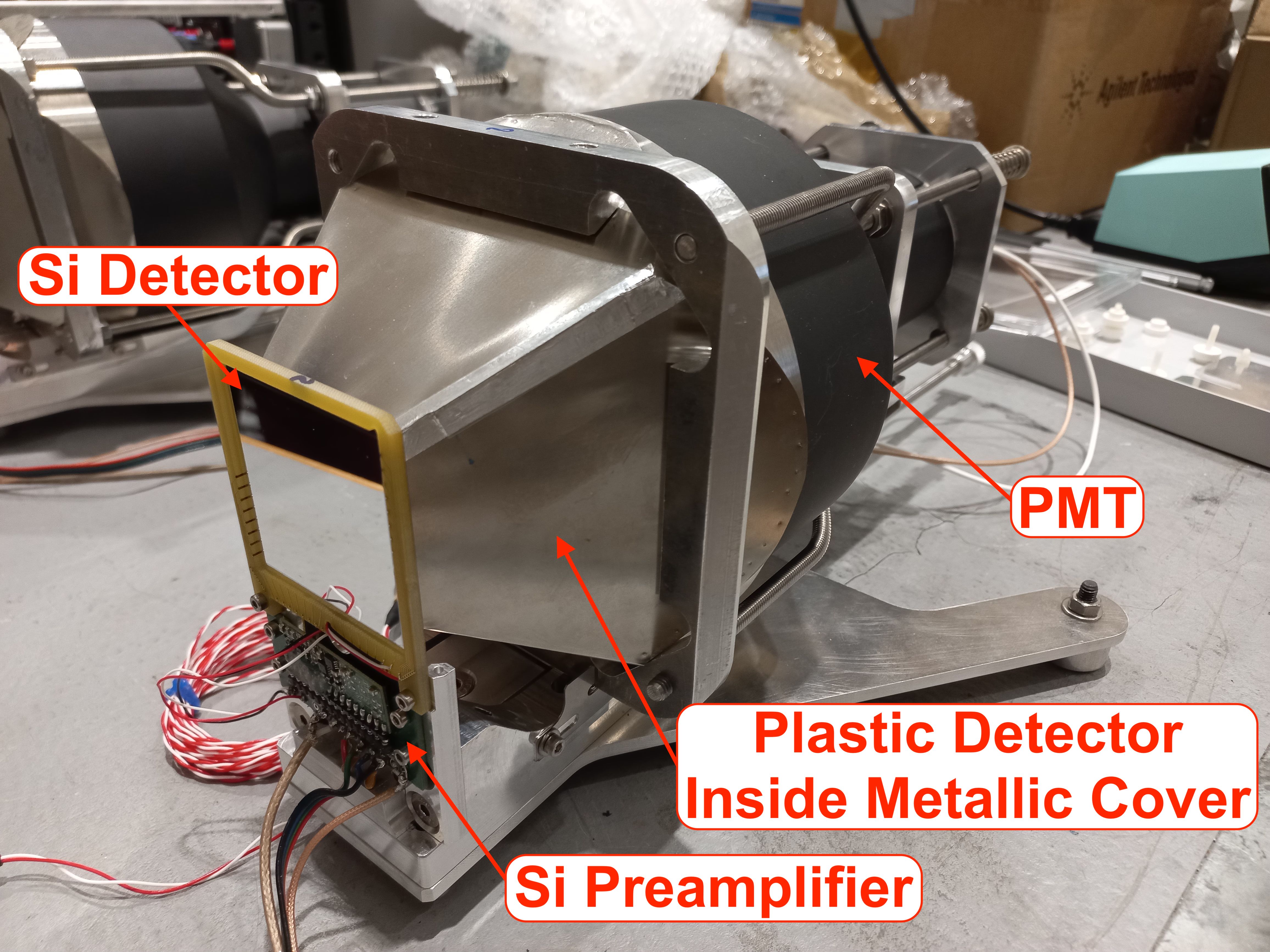}
		    \caption{Mounted \textit{e-Shape} detector before its installation inside the setup chamber.}
		    \label{fig_det}
		\end{figure}

        The scintillation light from the plastic detectors was collected with R877-02 PMTs, distributed by Hamamatsu. To maximize light collection, the sides of the plastic detectors were painted with the EJ510 reflective paint distributed by Eljen Technology. The PMTs were covered with the E989-26 metallic shielding from Hamamatsu. The Si detectors have CAEN A1422H preamplifiers (90 mV/MeV amplification factor) integrated into their plastic frames. Si preamplifiers are designed to operate both under vacuum and at atmospheric pressure. In coincidence mode, the \textit{e-Shape} detectors have an electron and $\gamma$-ray detection efficiency of around 6\% and 0.06\%, respectively. During measurements, the resolution of the Si detectors (FWHM) was 24 keV at 0.715 MeV, and 110 keV at 1 MeV for the plastic detectors in coincidence mode. The energy thresholds of the Si and plastic detectors were of the order of 50 and 30 keV, respectively. Fig. \ref{fig_det} shows an \textit{e-Shape} detector in its mounted configuration.
        
        It is important to highlight that the detection efficiency for $\gamma$-rays, in the energy range of the electromagnetic de-excitations of relevant daughter isotopes for reactor spectra, is negligible in the Si detector due to its small thickness. Consequently, Si–plastic coincidence spectra are dominated by the detection of electrons. In essence, the experimental $\beta$ spectrum is constructed from events in the plastic scintillator when they are in coincidence with signals from the Si detector \cite{GUADILLA}.

        Because coincidences with the Si detector are required, the energy threshold of this detector must be kept as low as possible to prevent the loss of information in the Si detector and, consequently, deformations in the coincidence spectra. This can be understood by considering the energy distribution of electrons in a thin Si layer, which follows the Landau distribution \cite{LANDAU}. This distribution has a Gaussian-like shape centered at the most probable energy deposited by electrons of a certain initial kinetic energy (e.g., 207~keV for 5~MeV electrons traversing a 527~$\mu$m Si layer). It is characterized by a short (highly attenuated) low-energy exponential-like tail and a long (slightly attenuated) high-energy exponential-like tail (see Fig.~1 of \cite{LANDAU}). Therefore, by ensuring that the Si threshold remains below the energy where the low-energy tail of the Landau distribution begins, no signal in the plastic detector associated with a coincidence in the Si detector is lost in the plastic coincidence spectrum. Conversely, if the Si threshold is set too high, it can produce deformations in the plastic coincidence spectrum, complicating the analysis. In such cases, possible signals in the plastic detector would not be registered as coincidence events, as their corresponding Si signals would fall below the threshold and therefore not be processed.

    \section{I233 Experiment}

		The IGISOL facility at the Department of Physics of the University of Jyv\"askyl\"a (JYFL) \cite{AYSTO} was chosen as the place to measure the $\beta$ spectra of interest since it can produce isotopically pure radioactive beams of isotopes relevant to address the question of the RAA and the ``bump'' \cite{ALGORA1}. At this facility, a 30 MeV proton beam induces fission reactions in a natural uranium target, and the fission fragments are extracted with a helium gas cell system. The design permits the extraction of refractory elements \cite{ALGORA1}. The purification of the beam is performed in two steps. First, a dipole magnet separates the fission fragments by their mass-to-charge ratio. This magnet has a mass resolving power (MRP) of $\sim$500 at an ion beam energy of 30 keV. Then the double Penning trap JYFLTRAP further purifies the beam \cite{ERONEN}. JYFLTRAP removes impurities from the radioactive beams by precisely controlling the motion of ions with magnetic and quadrupolar electric fields. For the isotopes measured in this experiment, only the first section of the trap, referred to as the ``purification trap'', was needed. The MRP of this trap is $\sim$10$^5$. The separation process takes up to a few ms, only inside the trap.

\begin{figure}[h]
		    \centering
		    \includegraphics[width=0.475\textwidth]{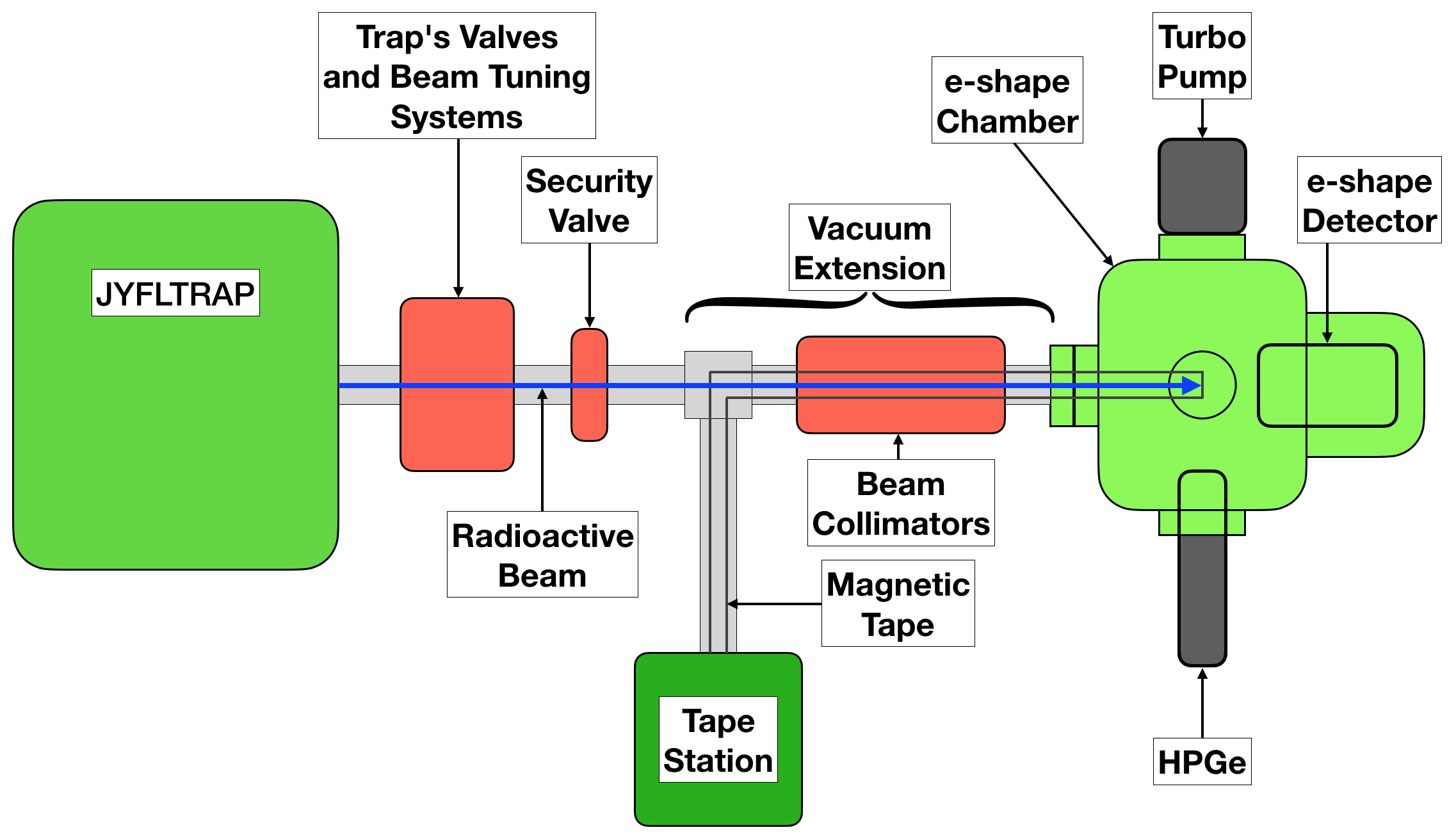}
		    \caption{Scheme of the I233 experimental setup (side view). The radioactive beam (blue arrow) is extracted from JYFLTRAP (left side of the figure) and passes through the trap's beam tuning system and security valve to the I233 setup (right side of the figure). A magnetic tape is drawn from the tape station, located below the beam trajectory. The tape is guided to the chamber where the detectors are placed for beam implantation. A vacuum extension tube connects the trap to the setup. A beam collimator system is installed in the vacuum extension tube. The \textit{e-Shape} chamber is connected to the vacuum extension and contains the \textit{e-Shape} detectors inside. The HPGe detector and a turbo pump are placed below and above the chamber, respectively.
}
		    \label{fig_I233SetUp}
		\end{figure}

        As explained above, the \textit{e-Shape} setup consists of two \textit{e-Shape} detectors \cite{GUADILLA}, each composed of a Si detector and a plastic detector. A HPGe detector and a CeBr$_3$ scintillator were added to monitor the beam. High-resolution $\gamma$-ray measurements enable \textit{in situ} identification of possible contaminants in the radioactive beam during initial beam tuning to the \textit{e-Shape} detector setup. This identification helps to improve the beam tuning and also supports the off-line analysis. The telescopes are installed inside a stainless-steel chamber designed to carry out measurements under vacuum. The HPGe and CeBr3 detectors are placed outside the chamber in front of thin Aluminium windows. The radioactive beams are guided from JYFLTRAP into the chamber via a system of collimators placed outside the chamber and at its entrance. The beams are implanted on a movable tape positioned in front of the \textit{e-Shape} detectors. Fig. \ref{fig_I233SetUp} shows a scheme of the I233 experiment setup used in the IGISOL facility.

        The signals from the detectors are processed with FASTER \cite{FASTER}, a triggerless modular digital data acquisition (DAQ) system developed at LPC Caen. FASTER includes various data processing algorithms to handle signals from many types of detectors. It timestamps the signals and has an automatic pile-up and saturation rejection system. Depending on the characteristics of the signals of a detector, the CARAS or MOSAHR daughterboards can be employed to digitize and process them. For the I233 experiment, the CARAS was used to process the signals from the plastic and CeBr3 detectors with QDC and TDC (charge-to-digital and time-to-digital converters) filters. The signals from the Si detectors were processed with both the CARAS and MOSAHR and the signals from the HPGe detector were processed with the MOSAHR, both daughterboards using an ADC (analog-to-digital converter) filter for these types of detectors.% A constant fraction discriminator (CFD) was employed to process the time information of the signals.

	\section{Spectra Analysis}

        To properly prepare the experimental data for analysis, it is necessary to identify both singles and coincidence events in the plastic detectors. Therefore, a suitable event time window must be chosen between the Si and plastic signals. Individual DAQ channel events flagged with FASTER pile-up and saturation warnings must be excluded from the experimental $\beta$ spectra to avoid introducing non-physical distortions in their shapes. It is worth noting that events registered with the DAQ and marked with the FASTER pile-up flag represent less than 2\% of all events recorded. Pile-up calculations with an independent algorithm were in agreement with the FASTER pile-up. Coincidences were reconstructed from the individual events. In this process, it is also necessary to exclude physical events in which multiple energies are deposited in the same detector within the event time window. This is a pile-up–like effect that increases as the event time window gets wider. Previous to the final processing of the events (event reconstruction), the length of the event reconstruction window was optimized. The goal was to reduce the event losses and avoid pile-ups in the reconstructed coincidence events. From this optimization, an event time window of 20~$\mu$s was found to be an acceptable value for the analysis of the experiment.

        Energy and resolution (width) calibrations were derived from features observed in the measured spectra. The maximum end-point DAQ channels, corresponding to the $Q$ values of the measured $\beta$ spectra, together with observed conversion electron peaks, were used for the calibrations. These calibration points relied on $Q$ values reported in the ENSDF database \cite{ENSDF}, and on conversion electron emission energies and intensities calculated with the BrIcc code \cite{KIBEDI_2008}. The endpoints used covered an energy range from 1.440 to 8.095 MeV.  

        Monte Carlo simulations, performed with the Geant4 code \cite{AGOSTINELLI_2003}, were employed to compute the response matrix of each \textit{e-Shape} detector. These detector responses are necessary to perform the deconvolution of the experimental spectra to obtain the real $\beta$ spectra \cite{TAIN} (see section \ref{sec_decv}). The Monte Carlo model of the experiment must first be validated to ensure its accuracy. The validation is achieved when simulated spectra suitably reproduce experimental measurements of well known sources. 

        To perform reliable deconvolutions, measured spectra must be free from background and contamination contributions. The corresponding measured background spectra were subtracted from the experimental data, and when necessary, simulated $\beta$ spectra of identified contaminants were also subtracted. Contributions of contaminations from $\beta$ decays of daughter nuclei were determined using the Bateman equations. In addition, HPGe detector spectra were used to identify $\gamma$-rays from contaminating sources and quantify their contributions to the $\beta$ spectra of interest.

        \subsection{Calibrations}
        
        The experiment energy calibration was determined by fitting the maximum end-point channels of the measured spectra using the standard shape of allowed $\beta$ spectra, including the statistical factor and an approximation of the Fermi function \cite{Venkataramaiah_1985}. E0 transition peaks from the $^{98}$Nb and $^{96}$Y $\beta$ decays were fitted with a double Gaussian model to account for contributions from the K and L atomic shells. A channel-to-expected-energy calibration approach was employed. Expected energies were obtained from ENSDF and BrIcc \cite{KIBEDI_2008}.
        
\begin{figure}[h]
		    \centering
		    \includegraphics[width=0.475\textwidth]{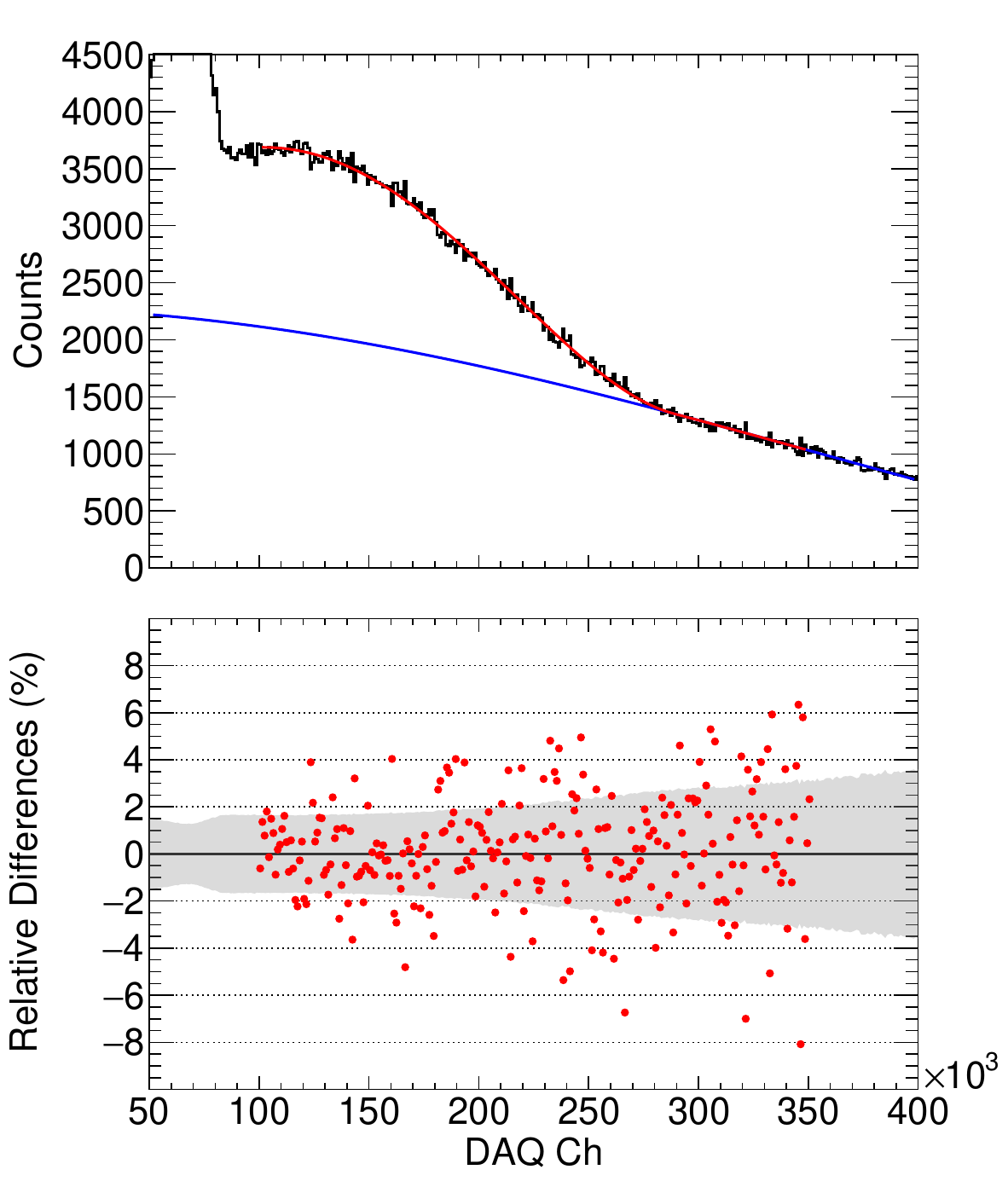}
		    \caption{Fit to the end-point channel of the $^{98}$Zr $\beta$ decay. Upper panel: The black line represents the plastic coincidence spectrum of the $^{98}$Zr$\rightarrow$$^{98}$Nb$\rightarrow$$^{98}$Mo $\beta$ decay chain measured with an \textit{e-Shape} detector. The blue line corresponds to the fit of the $^{98}$Nb $\beta$ spectrum, which serves as the background function for the total fit of the spectrum, represented by the red line, from which the end-point channel of $^{98}$Zr is obtained. Lower panel: The relative difference between the total fit and the experimental spectrum is shown, together with the one-sigma uncertainty band in gray. The relative differences are defined as $\frac{model-exp}{exp}$.}
		    \label{fig_enecalfit_ep}
		\end{figure}

\begin{figure}[h]
		    \centering
		    \includegraphics[width=0.475\textwidth]{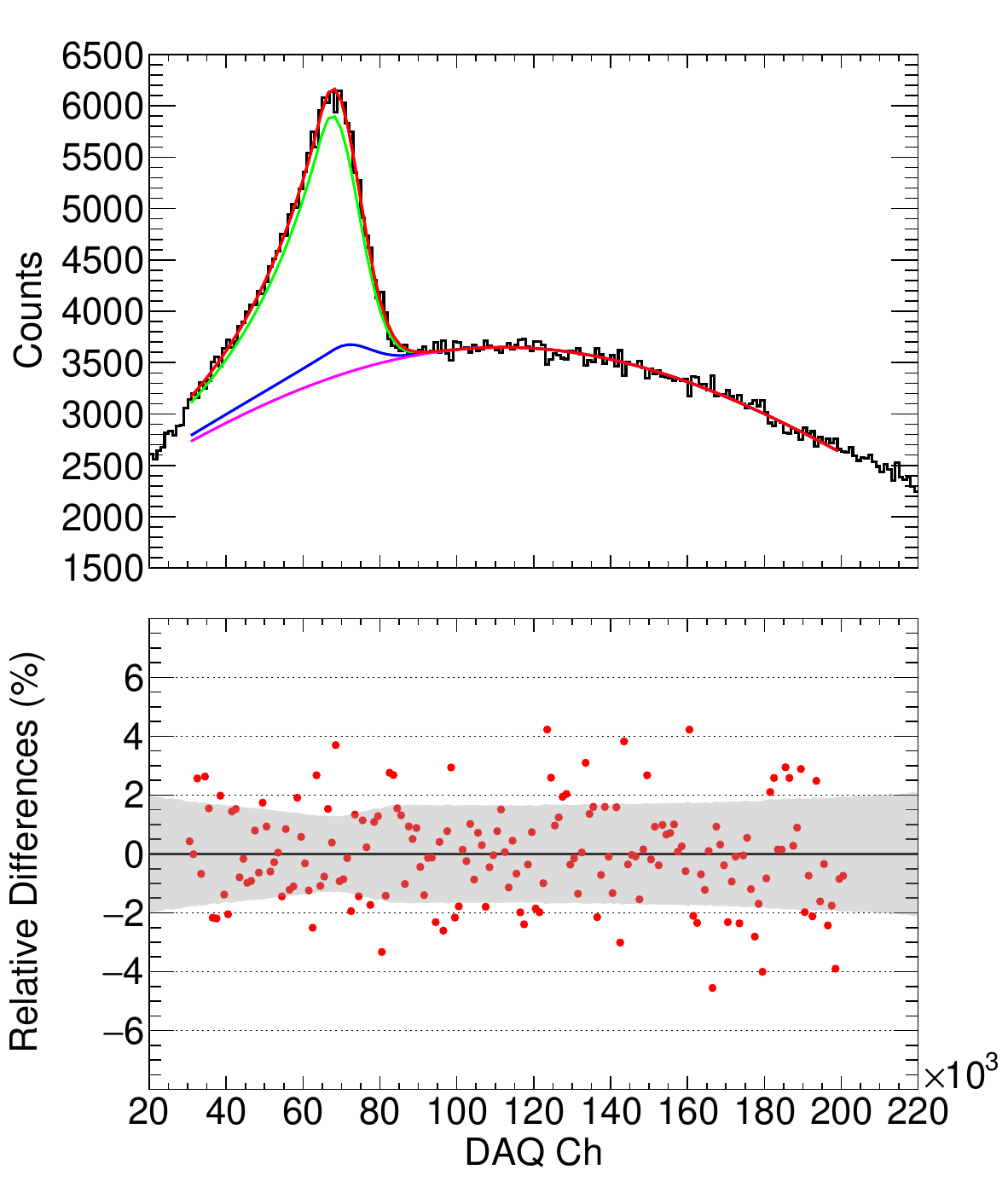}
		    \caption{Fit to the E0 conversion electron peak in the $^{98}$Nb $\beta$ decay. Upper panel: The black line represents the plastic coincidence spectrum of the $^{98}$Zr$\rightarrow$$^{98}$Nb$\rightarrow$$^{98}$Mo $\beta$ decay chain measured with an \textit{e-Shape} detector. The violet line corresponds to the fit of the $^{98}$Zr $\beta$ spectrum, which serves as the background function for the total fit of the spectrum, represented by the red line. The green and blue lines correspond to Gaussian fits with high- and low-energy exponential tails, associated with the conversion electrons emitted from the K and L1 atomic layers of $^{98}$Mo, respectively. The relative heights and channel separations correspond to the intensities and energy emissions provided by BrIcc \cite{KIBEDI_2008}, respectively. Lower panel: The relative difference between the total fit and the experimental spectrum is shown, together with the one-sigma uncertainty band in gray.
}
		    \label{fig_enecalfit_mulg}
		\end{figure}

        Fig.~\ref{fig_enecalfit_ep} shows an example of the fit to an end-point channel directly from a measured spectrum. The figure illustrates the fit to the $^{98}$Zr $\beta$ decay end-point channel, which corresponds to a $Q$ value energy of 2.238$\pm$0.010~MeV~\cite{HERZOG_1976}. The $^{98}$Zr$\rightarrow$$^{98}$Nb$\rightarrow$$^{98}$Mo $\beta$ decay chain was measured in this case, as it was not possible to separate the decay of $^{98}$Nb ($T_{1/2}$~=~2.86$\pm$0.06~s, $Q$~=~4.591$\pm$0.005~MeV) from that of $^{98}$Zr ($T_{1/2}$~=~30.7$\pm$0.4~s) because of their related decays and half-lives. Fig.~\ref{fig_enecalfit_mulg} shows an example of a fit to a conversion electron peak. The figure illustrates the double Gaussian fit to the conversion electron peak of the E0 transition in the $^{98}$Nb $\beta$ decay, accounting for the K and L1 atomic transitions ~\cite{KIBEDI_2008}.

		For the experimental resolution (or width) calibration, only two data points were available from the I233 experiment, the E0 transitions in $^{98}$Nb and $^{96}$Y at 0.715 and 1.564 MeV, respectively. This limited set of calibration data points was extended with the results of a test of the \textit{e-Shape} detectors performed with mono-energetic electrons at CENBG (Bordeaux, France) \cite{GUADILLA} that allowed us to characterize the plastic detectors better. 
        
        %%The combined data extended the range of the resolution calibration from 0.6 to 1.8 MeV. 

        %%The limited energy range of the resolution calibration may introduce small deviations in the energy calibration above 5~MeV. To reduce this effect, a high-energy correction was applied to refine the calibration. In this procedure, the expected energies of the data points in the high-energy region were slightly adjusted using a correction function derived from the differences between the preliminary $^{92}$Rb deconvolved spectrum, obtained with the uncorrected calibration, and a reference spectrum generated with reliable $\beta$ feedings and the standard allowed $\beta$ shape model.
        
        %After applying this correction, the agreement between experimental and simulated spectra was significantly improved, although a slight deformation around 6.3~MeV remains visible in some cases. This feature appears to depend on the intensity of the spectra in that region and may be related to the limited number of data points available for the energy and resolution calibrations. The width calibration, in particular, could only be extended up to 1.8~MeV using results from the CENBG test, while the spectra analyzed in this work reach up to 8~MeV. Future measurements will focus on increasing the number of calibration data points, and the analysis will be revisited to further improve the accuracy of the calibrations.

        \subsection{Monte Carlo Simulations}
        
        The response matrix of the \textit{e-Shape} detector was constructed with simulated spectra of monoenergetic electrons. Therefore, a previous validation of the Monte Carlo model was required to ensure that the simulation settings, such as the setup geometry and the physical interactions of simulated particles, accurately reproduced the experimental conditions. For this purpose, the $^{114}$Pd${\rightarrow}{}^{114}$Ag${\rightarrow}{}^{114}$Cd $\beta$ decay chain was measured to perform the validation. The decays of the $^{114}$Pd (T$_{1/2}$ = 2.42$\pm$0.06 min, $Q$ = 1.440$\pm$0.009 MeV) and $^{114}$Ag (T$_{1/2}$ = 4.6$\pm$0.1 s, $Q$ = 5.087$\pm$0.005 MeV) are dominated by allowed transitions, their $\beta$ feedings are known from high-resolution spectroscopy measurements \cite{ENSDF_114Ag}, and their half-lives ensure a one-to-one decay ratio. The detailed knowledge of these decays enables a reliable comparison between simulated and experimental spectra by applying only allowed shape corrections to the simulated spectra. Potential shape discrepancies arising from inaccuracies in the decay models are expected to be minimal.

\begin{figure}[h]
		    \centering
		    \includegraphics[width=0.475\textwidth]{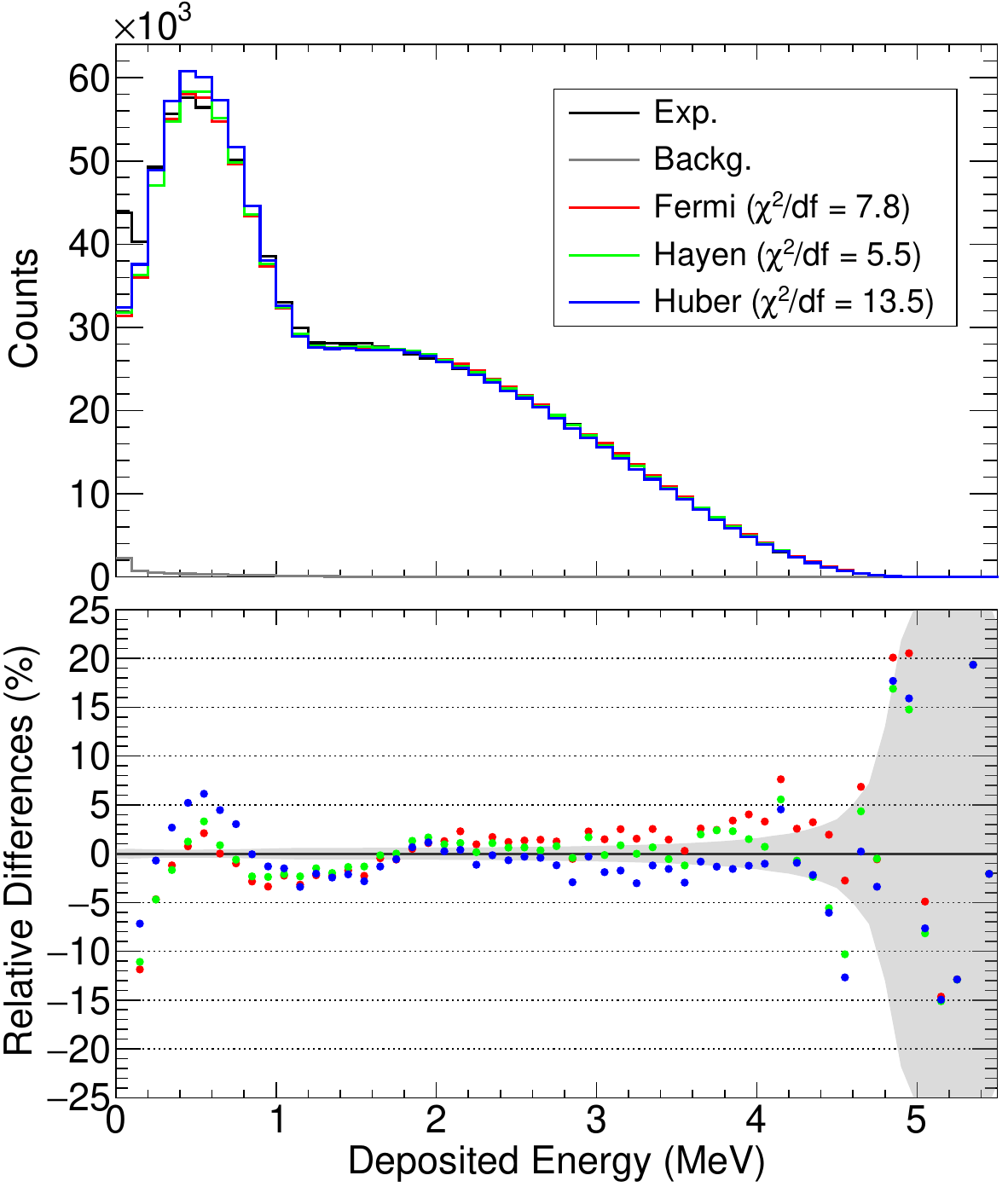}
		    \caption{Upper panel: Comparison of the measured spectrum of the combined $^{114}$Pd$\rightarrow$$^{114}$Ag$\rightarrow$$^{114}$Cd $\beta$ decays with Monte Carlo simulations using different shape models. The background (Backg.) is added to the simulations. Fermi refers to the basic $\beta$ shape with the Fermi function. Hayen and Huber denote the basic $\beta$ shape with the corrections of Hayen \textit{et al.} \cite{HAYEN_Allowed} and Huber \cite{HUBER}, respectively. The weak magnetism term of Huber \cite{HUBER} is included in the comparison with the corrections of Hayen \textit{et al.} \cite{HAYEN_Allowed}. Lower panel: Relative differences between the experimental and simulated spectra. A one-sigma error band is shown in grey in the lower panel.}
		    \label{fig_mcval}
		\end{figure}

		Simulations used the Geant4 Livermore modular physics list \cite{AGOSTINELLI_2003}. The geometry of the setup was constructed with precise Computer-Aided Design (CAD) models and integrated into the code with the CADMesh library \cite{CADMESH1,CADMESH2}. ENSDF $\beta$ feedings \cite{ENSDF_114Ag} for both nuclei were employed. Note that TAGS data are not available for the $\beta$ decay of both $^{114}$Pd and $^{114}$Ag. Three $\beta$ shape models were tested, a standard shape accounting only for the statistical factor and the Fermi function (Fermi), and two more including the most relevant allowed shape corrections of Hayen et al. \cite{HAYEN_Allowed} (Hayen) and Huber \cite{HUBER} (Huber), both with the Weak Magnetism term. Fig. \ref{fig_mcval} shows the plastic coincidence $^{114}$Pd-$^{114}$Ag experimental $\beta$ spectrum (Exp.) compared with the simulated spectra generated with the $\beta$ feedings and shape models outlined above. The experimental background spectrum (Backg.) is added to the simulations for comparisons. The $\chi^2$/df (chi-squared per degrees of freedom) values of the simulated spectra with respect to the experiment are given.

		The relative differences plot in Fig.~\ref{fig_mcval} shows that the simulation generated with the corrections of Hayen \textit{et al.} \cite{HAYEN_Allowed} remains below the 3\% level in most of the range of the experimental spectrum. The results of the other two simulations are comparable, with relative differences below 6\% to 3\%. These results fulfill our criterion for the validation of the Monte Carlo model for constructing the response matrix and performing deconvolutions, since the relative differences remain within the quality-threshold established in this analysis. This threshold was set on the order of a few percent ($\sim3\%$) to expect a reliable configuration of the simulation code.

        Due to the lack of previous experience of our collaboration in the analysis of $\beta$ spectra obtained with the \textit{e-Shape} detectors, we chose this quality-threshold criterion as appropriate. It represents an acceptable order for the relative differences to ensure the quality of reproduction of \textit{a priori} known experimental spectra with simulations, considering limitations related to the Monte Carlo code, uncertainties in the nuclear data and models required for the simulations, and possible uncertainties in the experimental data arising from detector effects and the presence of unidentified contaminants. Guaranteeing a validation with relative differences at the lower edge of the percentage level may allow us to start discerning general differences coming from the combined effect of the individual shape corrections included in the sets of corrections tested in this work (i.e., standard shape, Huber, and Hayen).
        
        To exemplify some of the commented limitations, it is worth noting that the discrepancies between the simulated spectra and the experimental data observed at low energies in the validation may arise from the use of $\beta$ feedings derived from high-resolution experiments \cite{ENSDF_92Rb,ENSDF_142Cs}, which could be slightly affected by the Pandemonium effect. Moreover, the loss of detection efficiency for low-energy electrons due to the presence of the Si detector can also influence the outcome of the simulations in this energy range.

        \subsection{Evaluation of Contaminants}
        
        Due to the comparable half-lives of the $\beta$ decays of the parent and daughter nuclei studied, the $\beta$ spectra of the parent nuclei can be contaminated by the decays of their daughters. To reduce these contaminations, tape moving cycles were implemented. At the end of each measuring cycle, the tape was moved to reduce the impact of the activity of the daughter nuclei. Tape moving cycles for $^{92}$Rb (T$_{1/2} = 4.48 \pm 0.03$~s, $Q = 8.095 \pm 0.006$~MeV) and $^{142}$Cs (T$_{1/2} = 1.684 \pm 0.014$~s, $Q = 7.308 \pm 0.011$~MeV) were set to 22.49 s and 7.15 s, respectively. Values reported for T$_{1/2}$ and $Q$ are taken from ENSDF \cite{ENSDF}.

		For the $^{92}$Rb and $^{142}$Cs cases, the daughter isotopes considered were $^{92}$Sr (T$_{1/2} = 2.611 \pm 0.017$~h, $Q = 1.951 \pm 0.009$~MeV) for the former, and $^{142}$Ba (T$_{1/2} = 10.6 \pm 0.2$~min, $Q = 2.212 \pm 0.005$~MeV) and $^{142}$La (T$_{1/2} = 91.1 \pm 0.5$~min, $Q = 4.504 \pm 0.005$~MeV) for the latter. The predicted contamination proportions calculated with Bateman equations for the corresponding tape moving cycles were 0.067\% for $^{92}$Sr, 0.31\% for $^{142}$Ba, and 0.0001\% for $^{142}$La. The experimental contamination proportions deduced from the respective HPGe spectra analyses were 1.92\% for $^{92}$Sr, 0.87\% for $^{142}$Ba, and 1.64\% for $^{142}$La.
		
		The experimental contamination proportions determined from the HPGe spectra of $^{92}$Sr and $^{142}$Ba were used as upper limits for cleaning the respective $^{92}$Rb and $^{142}$Cs experimental spectra, as these values are expected to account for undetermined contaminations present during the collection of the data. On the other hand, since $\gamma$-rays from $^{142}$La $\beta$ decays were observed in both the $^{142}$Cs measurement and background spectra, the predicted contamination proportion calculated with Bateman equations for $^{142}$La was used in the $^{142}$Cs spectrum analysis. The presence of $\gamma$-rays from $^{142}$La $\beta$ decays in the background spectra indicates that they are likely to originate from decays of $^{142}$Cs isotopes implanted at the entrance of the detector chamber. These $\gamma$-rays reach the HPGe detector, while the associated $\beta$ electrons are blocked by the thick aluminium collimator placed at the entrance of the chamber.

        There is also a small contribution from the natural background in the coincidence plastic spectra that must be subtracted. The natural background for the cases of interest was determined from experimental measurements conducted during the data taking of the corresponding cases to ensure a proper determination of these contamination spectra. For the $^{114}$Pd-$^{114}$Ag, $^{92}$Rb, and $^{142}$Cs cases, the natural background contamination proportions were 0.75\%, 0.66\%, and 1.93\% in the energy range of the decays.

		\subsection{Deconvolution} \label{sec_decv}
        
        Due to the response of the \textit{e-Shape} detectors, the measured experimental $\beta$ spectra are deformed with respect to that emitted originally. This is related to the combined effect of the energy distribution of electrons in the plastic and their energy losses when traversing the tape and the Si detector. Because of the low mass of electrons compared with atoms, they can undergo backscattering deflections, and their kinetic energies can decrease due to Bremsstrahlung effects. Consequently, the spectra of monoenergetic electrons in the plastic detectors are characterized by a Gaussian-like distribution followed by a low-energy tail. As a result, the Expectation–Maximization deconvolution method \cite{TAIN} was used iteratively to recover the real energy distribution of electrons from the measurements. The deconvolution of the data requires solving an inverse problem \cite{TAIN}, using the calculated response matrix together with the clean $^{92}$Rb and $^{142}$Cs experimental spectra.

\begin{figure}[h]
            \centering
            \includegraphics[width=0.475\textwidth]{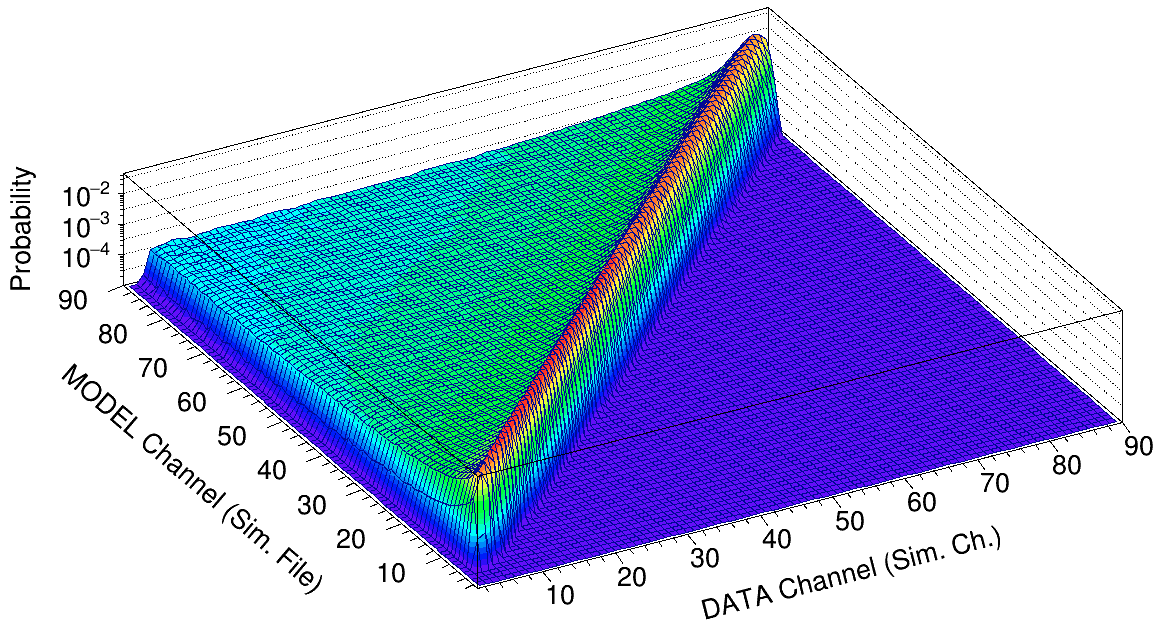}
            \caption{Three-dimensional representation of the response matrix of the plastic detector in coincidence, used for the analysis of the measured $\beta$ spectra. The y-axis represents the simulation files (or responses) of monoenergetic electrons for different initial kinetic energies. The x-axis represents the deposited energies (or channels of the simulated spectra) of the electrons in the plastic detector. The z-axis represents the normalized probability for monoenergetic electrons to deposit a given amount of energy in the plastic detector when detected in coincidence with the Si detector. In the terminology used for this analysis, the y-axis is referred to as the MODEL channel, as it represents the expected energies of the detected electrons, and the x-axis as the DATA channel, since it corresponds to the measured data.}
            \label{fig_resmatx}
        \end{figure}

        The inverse problem is based on finding a solution to the equation
        \begin{equation}
            \label{eq_Response}
            \vec{D} = \mathcal{R} \cdot \vec{O} + \vec{C}
        \end{equation}
        \noindent for $\vec{O}$, which represents the original (or true) $\beta$ spectrum. Here, $\vec{D}$ denotes the measured $\beta$ spectrum, $\vec{C}$ represents the contribution from contaminants in the measured spectrum, and $\mathcal{R}$ is the response matrix of the detector. Fig.~\ref{fig_resmatx} shows a three-dimensional representation of the response matrix for the plastic detector in coincidence, obtained from simulations of monoenergetic electrons and used in this work for the analysis of the experimental data. The response matrix was constructed from 450 simulations of monoenergetic electrons generated with initial kinetic energies ranging from 20 to 9000~keV in steps of 20~keV. For the deconvolutions presented in this work, after considering the resolution of the \textit{e-Shape} detectors, only the simulation files corresponding to energies that are multiples of 100~keV were used, and their simulated spectra were set with a bin width of 100~keV as well.

        For every iteration $j$ in the deconvolution process, the corresponding spectrum $\mathcal{R}\cdot\vec{O}_j$ is calculated (see Eq. \eqref{eq_Response}) and compared with $\vec D$. The iteration process stops when a target $\chi^2$ value between the $\mathcal{R}\cdot\vec{O}_j$ and experimental spectra is reached, or when a minimum number of iterations is completed. The deconvolution is considered successful when the calculated  $\mathcal{R}\cdot\vec{O}_j$ spectrum accurately reproduces the experimental spectrum across the entire energy region of the decay. It must be noted that contaminations in the experimental spectrum are subtracted before the start of the deconvolution process.

\begin{figure}[h]
		    \centering
		    \includegraphics[width=0.475\textwidth]{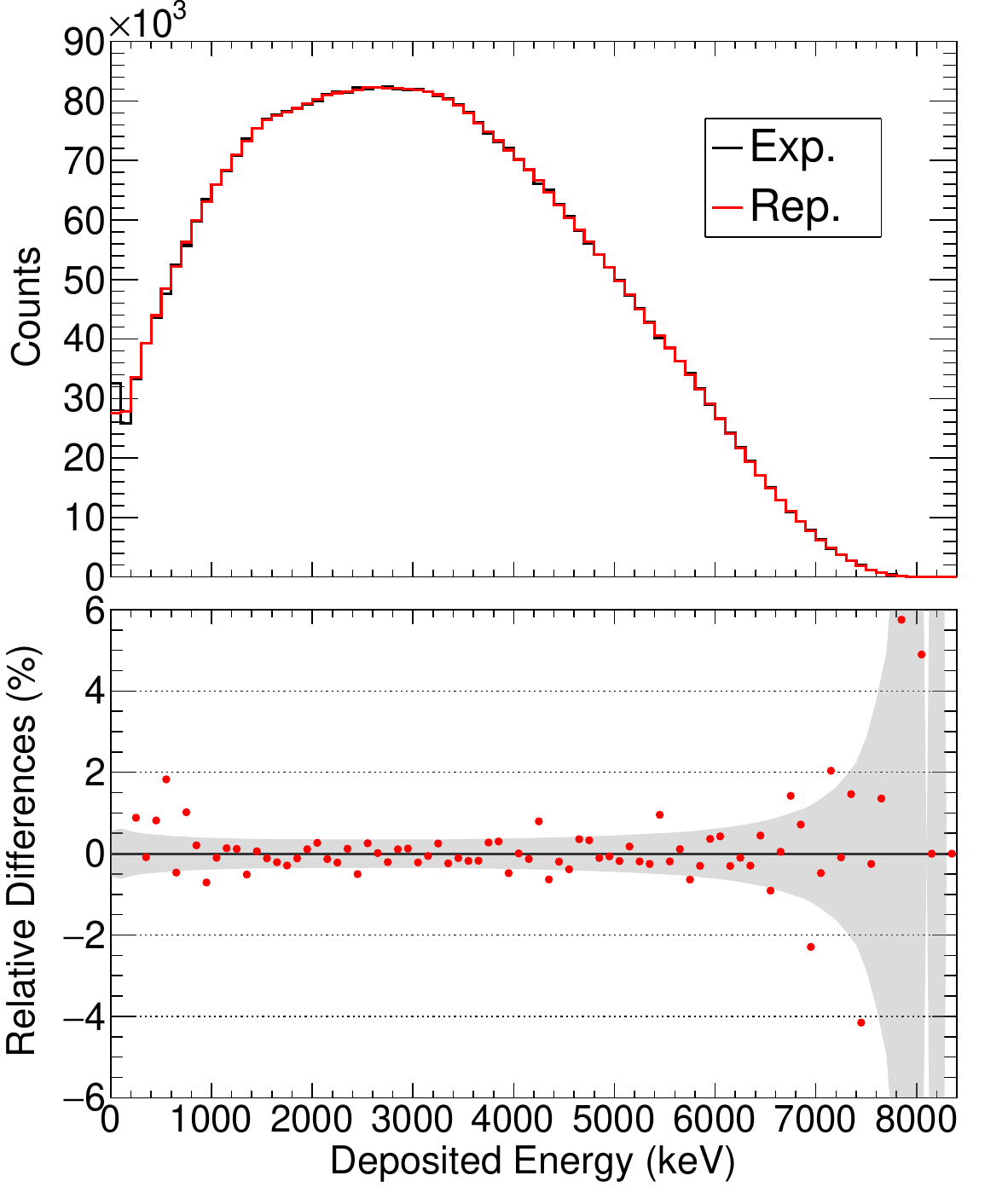}
		    \caption{Comparison of the measured $^{92}$Rb decay spectrum (Exp.), $\vec D$ in Eq. (\ref{eq_Response}), with the reproduction of the experiment obtained from the Expectation–Maximization deconvolution algorithm (Rep.), $\mathcal{R} \cdot \vec{O}$ in Eq. (\ref{eq_Response}). The relative differences are shown in the lower panel, with a one-sigma error band indicated in gray.}
		    \label{fig_repexp92Rb}
		\end{figure}

\begin{figure}[h]
		    \centering
		    \includegraphics[width=0.475\textwidth]{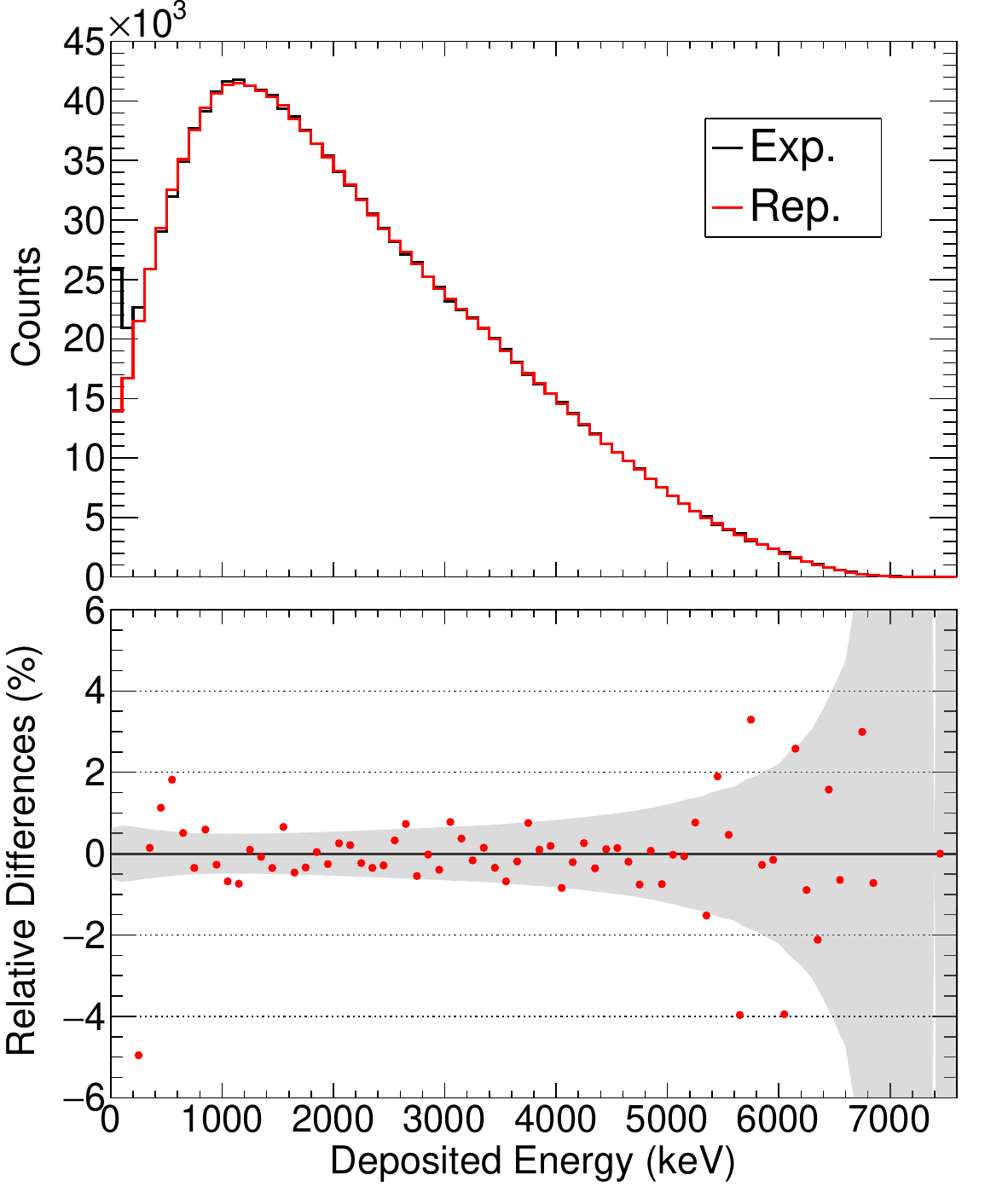}
		    \caption{Same as Fig. \ref{fig_repexp92Rb} but for $^{142}$Cs decay}
            %Comparison of the measured $^{142}$Cs decay spectrum (Exp.), $\vec D$ in Eq. (\ref{eq_Response}), with the reproduction of the experiment obtained from the Expectation–Maximization deconvolution algorithm (Rep.), $\mathcal{R} \cdot \vec{O}$ in Eq. (\ref{eq_Response}). The relative differences are shown in the lower panel, with a one-sigma error band indicated in gray.}
		    \label{fig_repexp142Cs}
		\end{figure}

        To demonstrate the effectiveness of the deconvolution algorithm used in this work for determining the original $\beta$ spectra of the $^{92}$Rb and $^{142}$Cs decays, the final reproduction of the $^{92}$Rb and $^{142}$Cs measured spectra are shown in Fig.~\ref{fig_repexp92Rb} and Fig.~\ref{fig_repexp142Cs}. As seen in the figures, the calculated spectrum $\mathcal{R}\cdot\vec{O}_{\text{final}}$ (Rep.) satisfactorily reproduces the experimental data (Exp.) across the entire energy range of the decay, indicating that the deconvolution process was carried out successfully and that the corresponding deconvolved spectrum ($\vec O$) is reliable.

        It should be noted that the tools and procedures used in this work to obtain the original $\beta$ spectra of interest share many similarities with those employed in the deconvolution of TAGS spectra. Therefore, the analysis of the $\beta$ spectra is also supported by the large experience gained by our collaboration in the analyses of many TAGS experiments.
        
        \subsection{Uncertainties}
        
        The following sources of uncertainty were considered and summed in quadrature for the deconvolved spectra obtained in this work: the statistical uncertainty of the counts per bin, the variance of the corresponding covariance matrix calculated from the analysis, and the systematic uncertainty related to the selection of the contaminant background. For the latter, two sets of deconvolved spectra were generated: one using the contribution of the experimentally determined contaminant (the results presented in this work) and another using the theoretically expected contribution. The relative difference between these two deconvolved spectra was calculated and added to the other uncertainties.
        
        The general contributions of these uncertainties are case dependent. Nonetheless, the values of the spectrum-weighted average per all bins for the statistical and covariance matrix uncertainties were approximately 0.13\% and 0.22\%, respectively. For the uncertainty related to the selection of the contaminant background, the corresponding average values were 3.0\% and 0.79\% for the $^{92}$Rb and $^{142}$Cs spectra, respectively.

	\section{Results}

        Before giving the extended results presented in our Letter \cite{ALCALA_PRL}, we briefly recall the main steps of the data analysis. The first step in obtaining the deconvolved spectra was the subtraction of contaminants from the measured spectra. In particular, the experimental $^{92}$Rb spectrum was cleaned by subtracting the corresponding background spectrum measured during the experiment, as well as a simulated $^{92}$Sr spectrum generated with ENSDF $\beta$ feedings \cite{ENSDF_92Rb} using the shape corrections of Hayen \textit{et al.} \cite{HAYEN_Allowed}. The simulated $^{92}$Sr spectrum was properly scaled to the experimental data using the contamination proportion determined from the $\gamma$ rays observed in the HPGe spectrum.

\begin{figure}[h]
		    \centering
		    \includegraphics[width=0.475\textwidth]{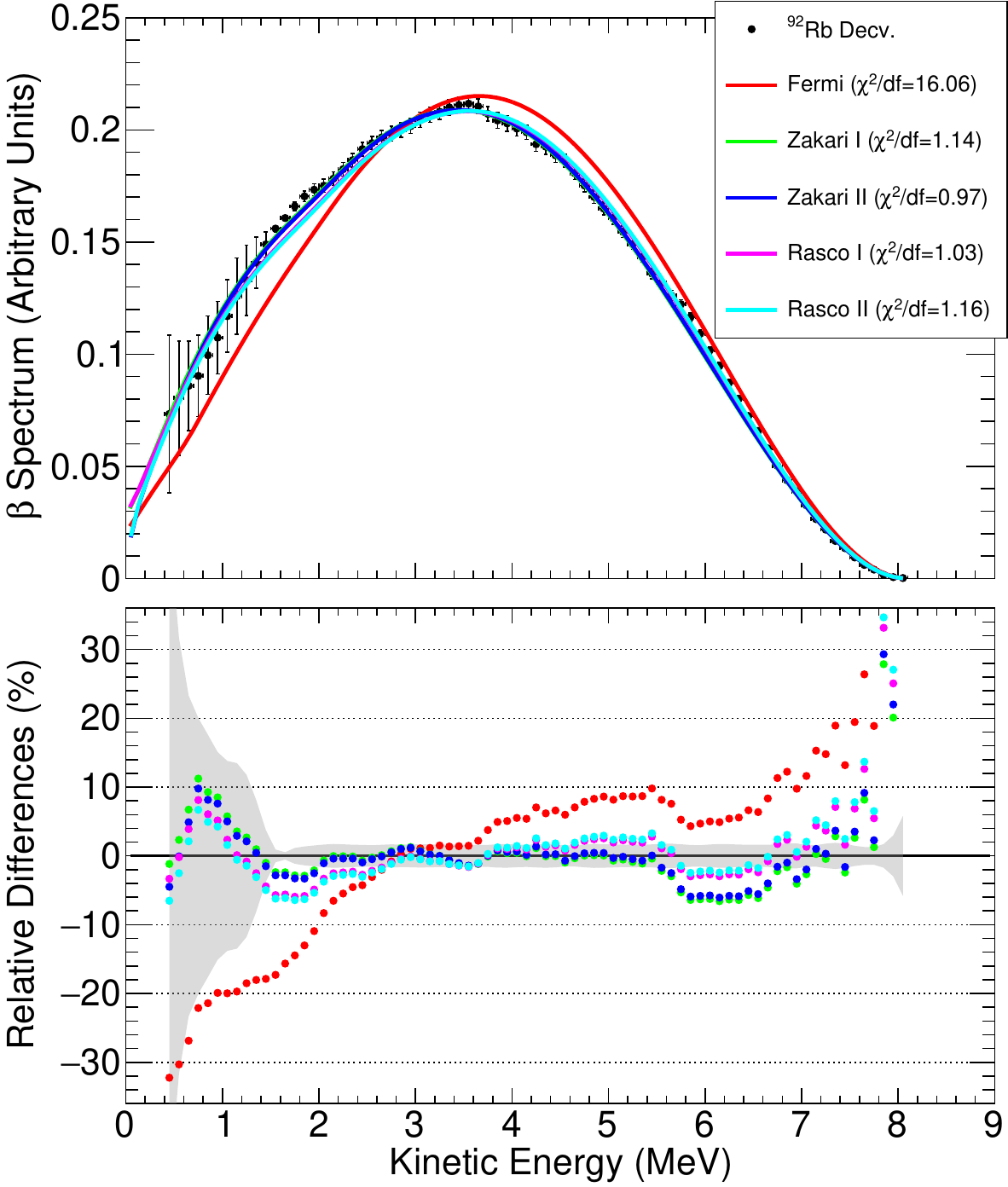}
		    \caption{Comparison of the deconvolved $\beta$ spectrum of the $^{92}$Rb $\rightarrow$ $^{92}$Sr decay ($^{92}$Rb Decv.) with various $\beta$ decay feedings and shape correction models. Fermi corresponds to the prediction obtained using ENSDF high-resolution feedings \cite{ENSDF_92Rb}, employing only the statistical factor and the Fermi function. Zakari I and II refer to predictions based on the TAGS feedings of Zakari-Issoufou \textit{et al.} \cite{ZAKARI}. For Zakari I, the allowed shape corrections of Hayen \textit{et al.} \cite{HAYEN_Allowed}, the weak magnetism term of Huber \cite{HUBER}, and the ground-state to ground-state first-forbidden shape correction of Hayen \textit{et al.} \cite{HAYEN_1F} were applied. For Zakari II, the allowed shape corrections of Huber \cite{HUBER} were used. Rasco I and II denote predictions based on the TAGS feedings of Rasco \textit{et al.} \cite{RASCO}, with the same shape corrections applied. The relative differences between the predicted and deconvolved spectra are presented in the lower panel. A one-sigma error band is shown in gray.}
		    \label{fig_decvRb}
		\end{figure}

		Fig. \ref{fig_decvRb} shows the deconvolved $^{92}$Rb spectrum ($^{92}$Rb Decv.), obtained with the Maximum-Entropy method \cite{TAIN}. This spectrum is compared with several predictions. A reference spectrum generated with ENSDF $\beta$ feedings \cite{ENSDF_92Rb} using only the statistical factor and the Fermi function (Fermi). Two spectra generated with TAGS $\beta$ feedings from Zakari-Issoufou \textit{et al.} \cite{ZAKARI}, employing the allowed shape corrections of Hayen \textit{et al.} \cite{HAYEN_Allowed} (Zakari I) and Huber \cite{HUBER} (Zakari II). Two spectra generated with TAGS $\beta$ feedings from Rasco \textit{et al.} \cite{RASCO}, employing the allowed shape corrections of Hayen \textit{et al.} \cite{HAYEN_Allowed} (Rasco I) and Huber \cite{HUBER} (Rasco II). The weak magnetism term is included in the Zakari I and Rasco I predictions. The ground-state to ground-state first-forbidden shape correction of Hayen \textit{et al.} \cite{HAYEN_1F} is only included in Zakari I and Rasco I.

		It is observed in Fig. \ref{fig_decvRb} that the shape difference between the deconvolved and reference (Fermi) spectra at low energies suggests that the ENSDF ground-state to ground-state $\beta$ feeding value of 95.2$\pm$0.7\% \cite{ENSDF_92Rb} is likely to be overestimated compared with the TAGS $\beta$ feeding values reported by Zakari-Issoufou \textit{et al.} \cite{ZAKARI} and Rasco \textit{et al.} \cite{RASCO}, 87.5$\pm$2.5\% and 91$\pm$3\%, respectively. Predictions generated with TAGS $\beta$ feedings (Zakari I and II, and Rasco I and II) show closer agreement with the deconvolved experimental data in the whole energy range of the decay when compared with the prediction generated with high-resolution $\beta$ feedings (Fermi), as indicated by their $\chi^2$/df values.

        It must be highlighted that the most relevant source of difference between the reference prediction (Fermi) and the predictions generated with TAGS $\beta$ feedings comes from the sets of feedings themselves. All predictions include the most important shape factors, the statistical factor, and the Fermi function. Consequently, differences among these $\beta$ spectra, related to the extra shape correction factors tested in this work, are below the percent level in the whole energy range of the decay.

		The $^{142}$Cs experimental spectrum was cleaned using the same procedure applied to the $^{92}$Rb data. In this case, $^{142}$Ba and $^{142}$La simulated spectra were generated with ENSDF $\beta$ feedings \cite{ENSDF_142Cs} and the shape corrections of Hayen \textit{et al.} \cite{HAYEN_Allowed}. The simulated spectra were scaled with the experimental contamination proportion for $^{142}$Ba and the predicted contamination proportion for $^{142}$La. The scaled simulated spectra were then subtracted from the experimental spectrum.

\begin{figure}[h]
		    \centering
		    \includegraphics[width=0.475\textwidth]{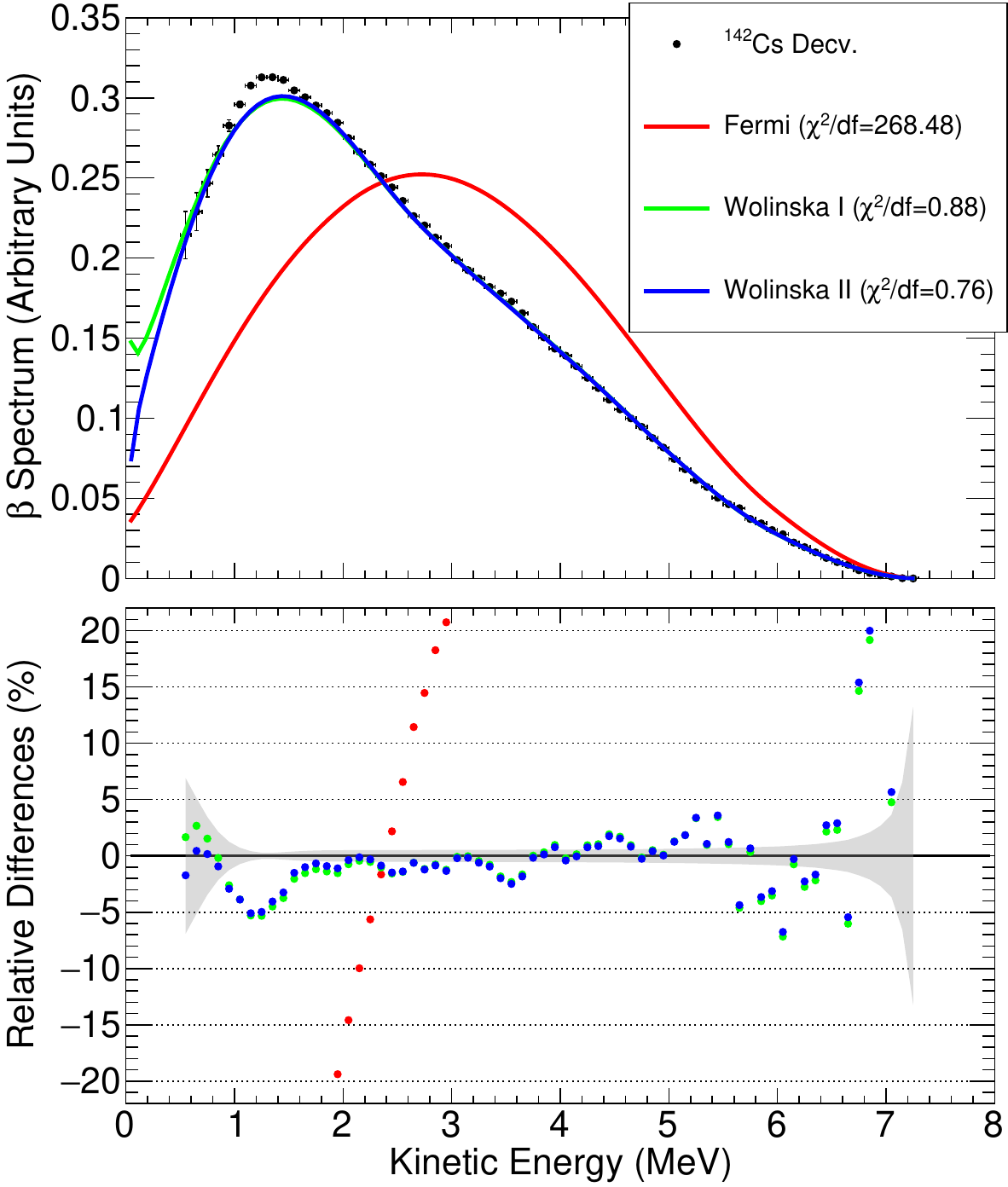}
		    \caption{Comparison of the deconvolved $\beta$ spectrum of the $^{142}\text{Cs}\rightarrow{}^{142}\text{Ba}$ decay ($^{142}$Cs Decv.) with various $\beta$-decay feeding and shape-correction models. Fermi corresponds to the prediction obtained using ENSDF high-resolution feedings \cite{ENSDF_142Cs}, employing only the statistical factor and the Fermi function. Wolinska I and II refer to predictions based on the TAGS feedings of Woli\'nska-Cichocka \textit{et al.} \cite{WOLINSKA}. For Wolinska I and II, the allowed shape corrections of Hayen \textit{et al.} \cite{HAYEN_Allowed} and Huber \cite{HUBER} were used, respectively. The weak magnetism term of Huber \cite{HUBER} was added to Wolinska I. The ground-state to ground-state first-forbidden shape correction of Hayen \textit{et al.} \cite{HAYEN_1F} was only applied to Wolinska I. The relative differences between the predicted and deconvolved spectra are presented in the lower panel. A one-sigma error band is shown in gray.}
		    \label{fig_decvCs}
		\end{figure}

		Fig. \ref{fig_decvCs} shows the deconvolved $^{142}$Cs spectrum ($^{142}$Cs Decv.), obtained with the Maximum-Entropy deconvolution method, compared with several predictions. These include a reference spectrum generated with ENSDF $\beta$ feedings \cite{ENSDF_142Cs} using only the statistical factor and the Fermi function (Fermi). And two spectra generated with TAGS $\beta$ feedings from Woli\'nska-Cichocka \textit{et al.} \cite{WOLINSKA}, incorporating the shape corrections of Hayen \textit{et al.} \cite{HAYEN_Allowed} (Wolinska I) and Huber \cite{HUBER} (Wolinska II). The weak magnetism term is included in the Wolinska I prediction. The ground-state to ground-state first-forbidden shape correction of Hayen \textit{et al.} \cite{HAYEN_1F} is only included in Wolinska I.

		Fig. \ref{fig_decvCs} shows significant differences between the deconvolved and reference (Fermi) spectra, suggesting that the $^{142}$Cs $\beta$ feedings reported in ENSDF \cite{ENSDF_142Cs} are likely to be affected by the Pandemonium effect \cite{ALGORA1}. The $\chi^2$/df values of these predictions and the relative differences show that the best description of the experimental data is obtained with the predictions generated with the TAGS $\beta$ feedings of Woli\'nska-Cichocka \textit{et al.} \cite{WOLINSKA}. 
        
        The only observable discrepancy between the $^{142}$Cs deconvolved data and the Wolinska I and II predictions is found around the 1.3 MeV energy region. This relative difference of 5\% between the predictions and the data may be caused by small inaccuracies in the $\beta$ feedings of Woli\'nska-Cichocka \textit{et al.} \cite{WOLINSKA}, which could be addressed by revisiting the strength of the most relevant feedings of their $\beta$ decay model. Another possible source for this discrepancy may be the lack of inclusion of appropriate first-forbidden shape correction factors for transitions with $\Delta I>0$, as this decay has several non-negligible feedings of transitions of those types.

		The deconvolved $\beta$ spectra were also compared with the corresponding spectra of Rudstam \textit{et al.} \cite{RUDSTAM_1990}. The latter were based on measurements by Tengblad \textit{et al.} \cite{TENGBLAD} conducted at OSIRIS and ISOLDE in the 1980s. 
        %Figs. \ref{fig_compRb} and \ref{fig_compCs} illustrate comparisons between the deconvolved $\beta$ spectra presented in this article and the corresponding spectra from Rudstam et al. \cite{RUDSTAM_1990} for the decays of $^{92}$Rb and $^{142}$Cs, respectively.

		\begin{figure}[h]
		    \centering
		    \includegraphics[width=0.475\textwidth]{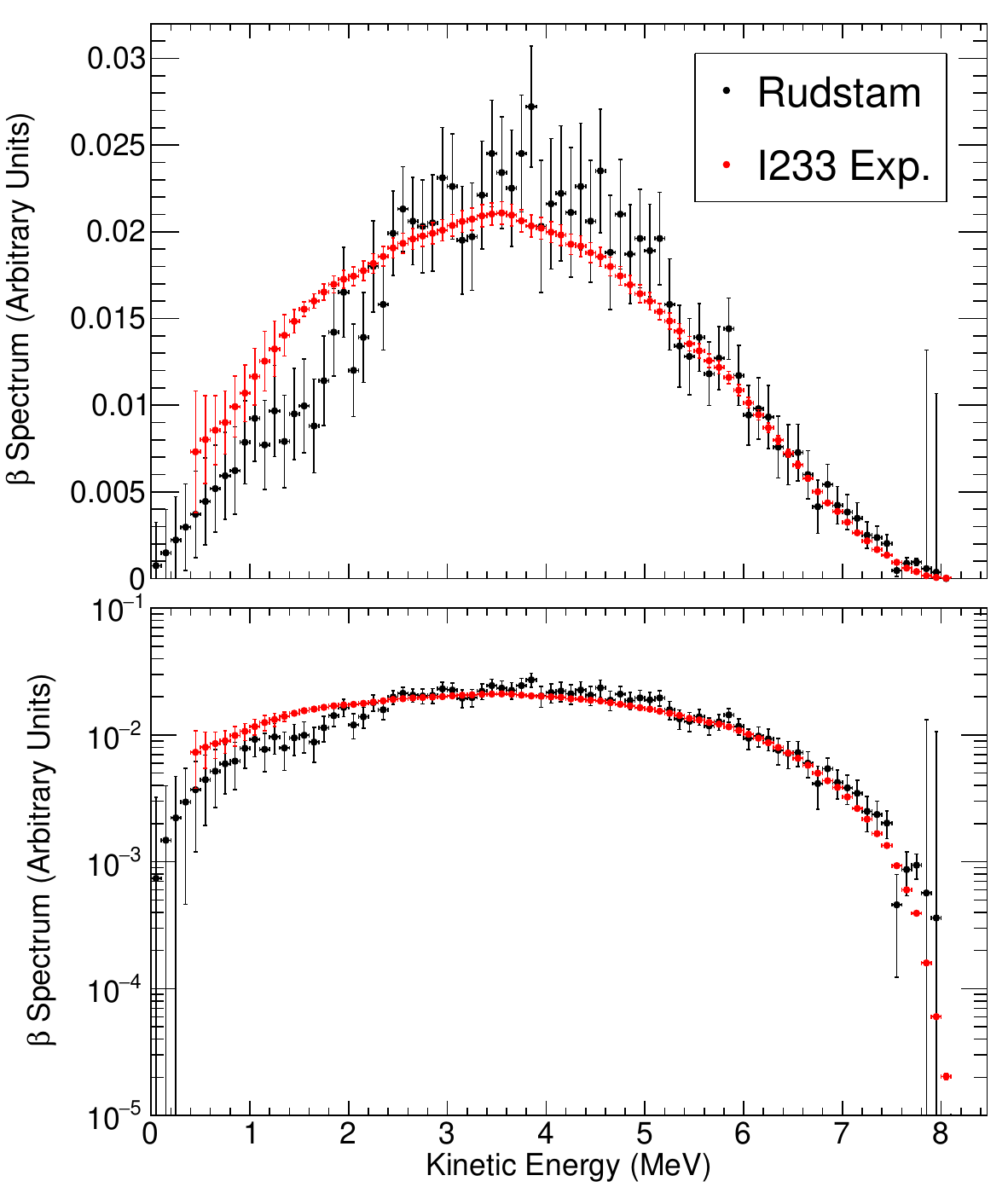}
		    \caption{Comparison of the $^{92}$Rb deconvolved $\beta$ spectrum (I233 Exp.) with the corresponding spectrum of Rudstam \textit{et al.} \cite{RUDSTAM_1990} (Rudstam). Linear and logarithmic plots are shown in the upper and lower panels, respectively.}
		    \label{fig_compRb}
		\end{figure}

        Fig. \ref{fig_compRb} compares the $^{92}$Rb $\beta$ spectrum obtained in this work (I233 Exp.) with the same spectrum reported by Rudstam \textit{et al.} \cite{RUDSTAM_1990}. The most significant differences are observed at energies below 2 MeV, where the data of Rudstam \textit{et al.} \cite{RUDSTAM_1990} exhibit a shape deformation compared to the deconvolved spectrum presented in this article. Additionally, other discrepancies are observed between 3 and 5 MeV, as well as near the $Q$ value energy of the decay. The differences at high energies are expected due to the intrinsically low statistics of $\beta$ spectra in the high-energy region. The experimental spectrum reported in the present article is smoother and has smaller uncertainties compared to that of Rudstam \textit{et al.} \cite{RUDSTAM_1990}.

		\begin{figure}[h]
		    \centering
		    \includegraphics[width=0.475\textwidth]{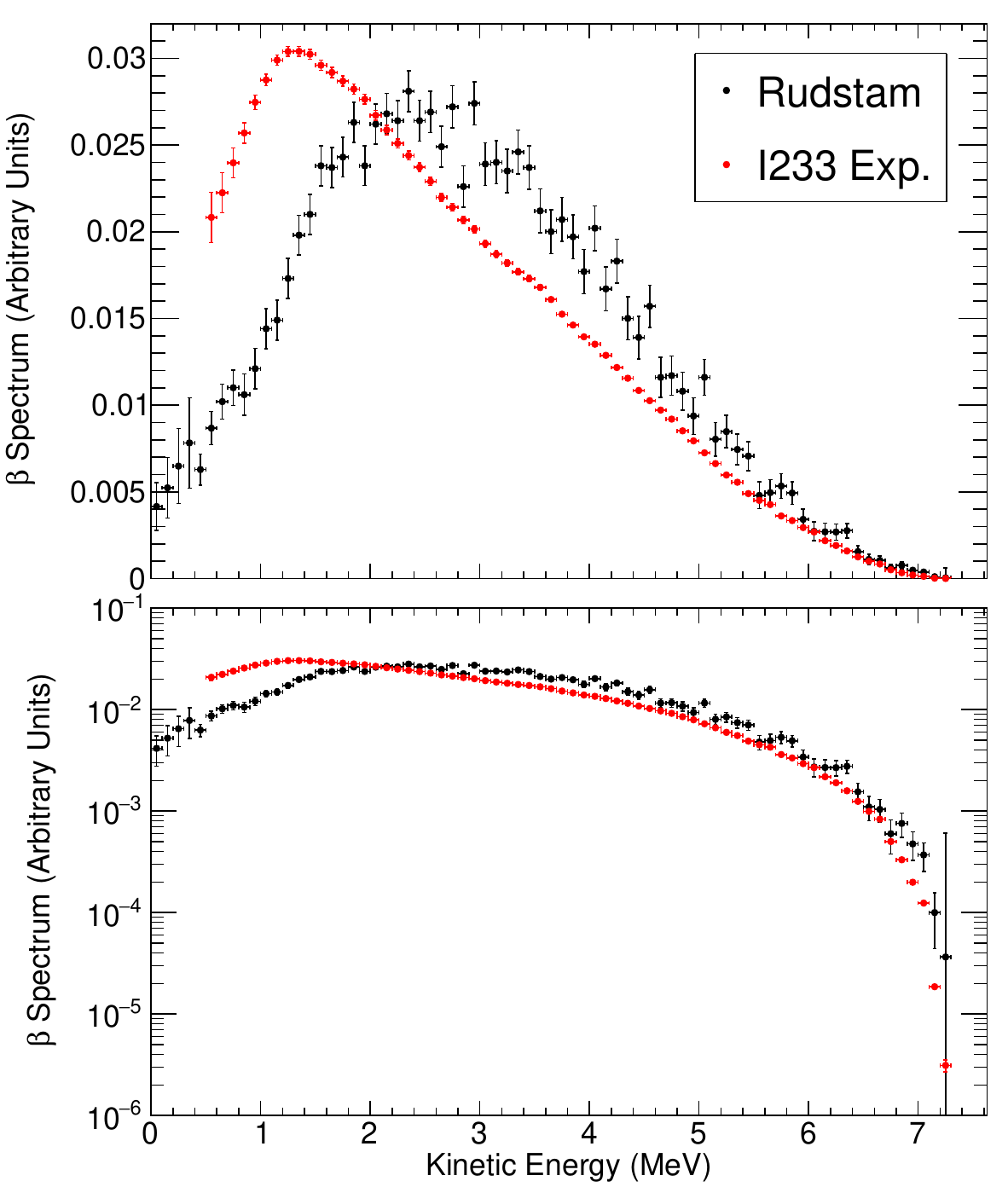}
		    \caption{Comparison of the $^{142}$Cs deconvolved $\beta$ spectrum (I233 Exp.) with the respective spectrum of Rudstam \textit{et al.} \cite{RUDSTAM_1990} (Rudstam). Linear and logarithmic plots are shown in the upper and lower panels, respectively.}
		    \label{fig_compCs}
		\end{figure}

        Fig. \ref{fig_compCs} compares the $^{142}$Cs $\beta$ spectrum obtained in this work (I233 Exp.) with the same spectrum reported by Rudstam \textit{et al.} \cite{RUDSTAM_1990}. There are significant differences at energies below 2 MeV. Most likely, as a consequence of the low-energy deformation, the deconvolved spectrum also differs from that of Rudstam \textit{et al.} \cite{RUDSTAM_1990} in the 2 to 6 MeV energy region. As in the $^{92}$Rb case, the differences at high energies are generally expected due to the intrinsically low statistics of $\beta$ spectra in their high-energy region. The experimental spectrum reported in this work is also smoother and has smaller uncertainties compared to that of Rudstam \textit{et al.}.

        \subsection{First-Forbidden Correction}
        
        The effects of forbidden correction factors on $\beta$ spectra shapes are highly complex. These corrections are highly dependent on the transition matrix elements connecting the nuclear states of the corresponding decays and require extensive calculations \cite{BEHRENS_1971,BEHRENS_BOOK}. Consequently, their shapes can vary considerably according to changes in the nuclear properties of the decaying nuclei, such as the change in nuclear spin and parity ($\Delta I^\pi$).

        Hayes \textit{et al.} \cite{HAYES2} provide a set of general shape correction factors for allowed and first-forbidden Gamow-Teller and Fermi $\beta$ decays, accounting for $\Delta I^\pi = 0^-, 1^-, 2^-$, that can be used as approximations in decays relevant to reactor spectra. Nonetheless, Hayen \textit{et al.} \cite{HAYEN_1F} calculated non-unique first-forbidden shape correction factors for specific $\beta$ transitions with significant contributions to reactor spectra. The transitions were selected according to their $\beta$ branching ratios and parent fission yields. Hayen \textit{et al.} \cite{HAYEN_1F} determined the first-forbidden shape correction factors of the $^{92}$Rb and $^{142}$Cs ground-state to ground-state $\beta$ transitions, which were implemented in this work. An essential aspect of these correction factors is that they are associated with a nuclear spin-parity change of $\Delta I^\pi = 0^-$, and such corrections are expected to affect the shape of the spectra minimally. Thus, they constitute a good test of the proper control of the responses of the \textit{e-Shape} detectors.
 
\begin{figure}[h]
		    \centering
		    \includegraphics[width=0.475\textwidth]{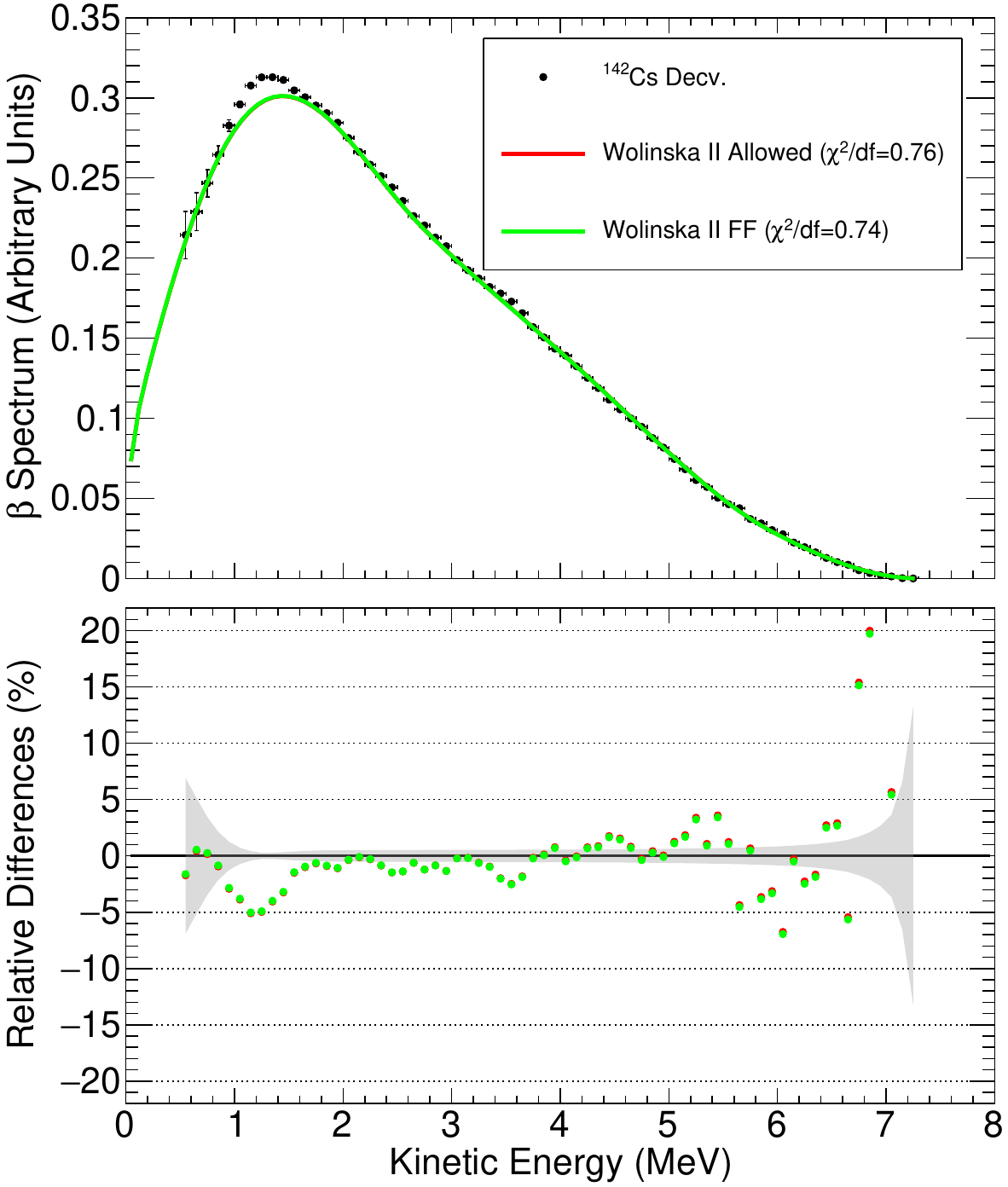}
		    \caption{Comparison of the deconvolved $^{142}$Cs $\beta$ spectrum ($^{142}$Cs~Decv.) with two model predictions based on the TAGS $\beta$ feedings of Woli\'nska-Cichocka \textit{et al.} \cite{WOLINSKA}. The predictions use the allowed shape corrections of Huber \cite{HUBER}, both without (Wolinska~II~Allowed) and with (Wolinska~II~FF) the ground-state to ground-state first-forbidden correction factor of Hayen \textit{et al.} \cite{HAYEN_1F}. The relative differences between the predicted and deconvolved spectra are presented in the lower panel. A one-sigma error band is shown in gray.}
		    \label{fig_decvRb_FFcor}
		\end{figure}

        To illustrate the effect of first-forbidden shape correction factors associated with $\Delta I^\pi = 0^-$ $\beta$ transitions, Fig.~\ref{fig_decvRb_FFcor} shows the deconvolved $^{142}$Cs spectrum ($^{142}$Cs~Decv.) compared with predictions generated with the TAGS $\beta$ feedings of Woli\'nska-Cichocka \textit{et al.} \cite{WOLINSKA}. The allowed shape corrections of Huber \cite{HUBER} were included, both without (Wolinska~II~Allowed) and with (Wolinska~II~FF) the ground-state to ground-state first-forbidden shape correction of Hayen \textit{et al.} \cite{HAYEN_1F}. The $^{142}$Cs $\beta$ spectrum is a representative case to test the effect of the ground-state to ground-state first-forbidden shape correction since this decay has a relatively large contribution of the ground-state to ground-state $\beta$ feeding, with a value of 40.6$\pm$2.0\% reported from TAGS measurements \cite{WOLINSKA}.

        Fig.~\ref{fig_decvRb_FFcor} shows that, for the $^{142}$Cs $\beta$ spectrum, the calculated ground-state to ground-state first-forbidden correction factor related to a $\Delta I^\pi=0^-$ $\beta$ transition \cite{HAYEN_1F} has a negligible effect in the whole energy region of this decay. The predictions generated only with allowed corrections (Wolinska II Allowed) and the one including the respective first-forbidden correction factor (Wolinska II FF) are almost indistinguishable from each other, and both are consistent with the shape of the deconvolved experimental spectrum. This experimentally confirms that the ground-state to ground-state branch of this decay does not have sensitivity to first-forbidden transition effects since it is of the $\Delta I^\pi=0^-$ type. This observation is also consistent with the analogous non-unique first-forbidden shape correction factor proposed by Hayes \textit{et al.} \cite{HAYES2}.

        \subsection{Ground-State Feeding Optimization}
        
        An important application of a measured $\beta$ spectrum is in its use to constrain or validate the value of the ground-state to ground-state feeding, or any other dominant feeding of the decay, obtained with different experimental feedings (e.g., a TAGS analysis). This constraining is performed by comparing the deconvolved spectrum with the respective predictions generated with the output of the feeding model. 
        %This approach provides an additional constraining method to determine ground state to ground state feedings to that presented in Guadilla \textit{et al.} \cite{GUADILLA3,GUADILLA_GS}.
        
        The cases studied were the TAGS ground-state to ground-state feedings of the $^{92}$Rb decay reported by Zakari-Issoufou \textit{et al.} \cite{ZAKARI} and Rasco \textit{et al.} \cite{RASCO}, as well as the corresponding feeding of the $^{142}$Cs decay reported by Woli\'nska-Cichocka \textit{et al.} \cite{WOLINSKA}. To perform the comparisons, different predicted $\beta$ spectra from the same initial model were generated by varying the value of the feeding of interest in steps of 0.1\%. For each test value of the modified feeding, the remaining feedings were proportionally adjusted to maintain the total normalization of all feedings to 100\%. %To better illustrate this procedure, for example, if a hypothetical initial ground-state feeding is 75\% and is reduced to 50\%, the 25\% difference subtracted from this ground-state feeding is redistributed equally among all other feedings in the set, ensuring that the total sum of all feedings remains 100\%.

        A $\chi^2$ test was implemented to determine the optimized ground-state to ground-state feedings of interest. For the $^{92}$Rb case, Zakari-Issoufou \textit{et al.} \cite{ZAKARI} reported a ground-state to ground-state feeding of 87.5$\pm$2.5\%. The optimized values obtained with the procedure described here are 89.0$\pm$2.6\% and 88.5$\pm$2.5\%, using the allowed corrections of Hayen \textit{et al.} \cite{HAYEN_Allowed} and those of Huber \cite{HUBER}, respectively. Both sets of corrections included the weak magnetism term of Huber \cite{HUBER} and the ground-state to ground-state first-forbidden correction of Hayen \textit{et al.} \cite{HAYEN_1F}. For the same decay, Rasco \textit{et al.} \cite{RASCO} reported a ground-state to ground-state feeding value of 91$\pm$3\%. The optimized values obtained are 91.1$\pm$2.1\% and 90.7$\pm$2.1\%, employing the same respective shape corrections. For the $^{142}$Cs case, Woli\'nska-Cichocka \textit{et al.} \cite{WOLINSKA} reported a value of 40.6$\pm$2.0\%. The optimized values determined in this work are 39.6$\pm$2.8\% and 39.5$\pm$2.8\%, using the respective sets of shape corrections used in the $^{92}$Rb case equivalent for the $^{142}$Cs decay. For completeness, the average values of the optimized ground-state to ground-state feedings weighted by their uncertainties are 90.0$\pm$1.1\% and 39.6$\pm$2.0\% for the $^{92}$Rb and $^{142}$Cs case, respectively. All optimized values are within the uncertainties of the corresponding reference feedings determined from the TAGS measurements. Table \ref{tab_optgs} contains the reported ground-state to ground-state $\beta$ feeding values for the $^{92}$Rb and $^{142}$Cs $\beta$ decays.

        \begin{table}[ht]
            \centering
            \caption{Comparison of the $^{92}$Rb and $^{142}$Cs ground-sate to ground-state (G.S. to G.S.) TAGS $\beta$ feeding values reported in the literature, against the average of the optimized values (Avg. Opt.) reported in our work.}
            \label{tab_optgs}
            \begin{tabular}{|c|c|}
                \hline
                \multicolumn{2}{|c|}{\textbf{$\boldsymbol{^{92}}$Rb G.S. to G.S. $\boldsymbol\beta$ feeding}} \\
                \hline
                Avg. Opt.                                     & 90.0$\pm$1.1\% \\
                \hline
                Zakari-Issoufou \textit{et al.} \cite{ZAKARI} & 87.5$\pm$2.5\% \\
                \hline
                Rasco \textit{et al.} \cite{RASCO}            & 91$\pm$3\%     \\
                \hline
                ENSDF \cite{ENSDF_92Rb}                       & 95.2$\pm$0.7\% \\
                \hline
                \multicolumn{2}{|c|}{\textbf{$\boldsymbol{^{142}}$Cs G.S. to G.S. $\boldsymbol\beta$ feeding}} \\
                \hline
                Avg. Opt.                                           & 39.6$\pm$2.0\% \\
                \hline
                Woli\'nska-Cichocka \textit{et al.} \cite{WOLINSKA} & 40.6$\pm$2.0\% \\
                \hline
                ENSDF \cite{ENSDF_142Cs}                            & 56$\pm$5\%     \\
                \hline
            \end{tabular}
        \end{table}

        \subsection{Average Energies}
        
        Due to its relevance in other areas of reactor physics research (e.g., decay heat calculations), the experimental average kinetic energy of the $^{92}$Rb and $^{142}$Cs $\beta$ decay electrons (or mean $\beta$ energy) was calculated from the corresponding deconvolved spectra using the formula
        \begin{equation}
            \label{eq_meanbetaene}
            \langle E_\beta \rangle = \frac{ \sum_{i=1}^{N} S_{\beta,i} \, E_{\beta,i} }{ \sum_{i=1}^{N} S_{\beta,i} } \;,
        \end{equation}
        \noindent which is valid for a discrete spectrum with equally spaced energy values, where $S_{\beta,i}$ is the $\beta$ spectrum value for a given kinetic energy $E_{\beta,i}$ of the $\beta$ electron, and $N$ is the total number of data points in the spectrum. For continuum predicted spectra, the version of the formula with integrals was used.

        As observed in Fig.~\ref{fig_decvRb} and Fig.~\ref{fig_decvCs}, there is a lack of deconvolved data points below 0.5~MeV due to the design of the \textit{e-Shape} detectors. $\beta$ electrons with kinetic energies below this threshold are difficult to detect in coincidence because of the average energy loss associated with their passage through the Si detector ($\Delta E$ part). This average energy loss is around 210~keV for the thickness of the Si detectors used \cite{LANDAU}. Because of this limitation, the model $\beta$ spectra corresponding to the smallest $\chi^2$/df values (see figures Fig.~\ref{fig_decvRb} and Fig.~\ref{fig_decvCs}) were normalized to the deconvolved experimental data. Model data points for the corresponding low-energy bins were then computed from the predictions and incorporated in the experimental data sets. This procedure was necessary to obtain the experimental average $\beta$ energy without the bias introduced by the absence of low-energy deconvolved data points. 
        
        For the $^{92}$Rb case, the best model prediction is given by the Zakari II spectrum (see Fig.~\ref{fig_decvRb}). After complementing the experimental points with the model calculation, the average kinetic energy for this decay is 3.52$\pm$0.03~MeV. For the $^{142}$Cs case, the best prediction corresponds to Wolinska II (see Fig.~\ref{fig_decvCs}), and the associated average kinetic energy for this decay is 2.41$\pm$0.01~MeV. For the $^{92}$Rb case, according to ~\cite{ALGORA1}, the average $\beta$ energies deduced from high-resolution and TAGS measurements are 3.640$\pm$0.030~MeV and 3.498$\pm$0.105~MeV, respectively. For the $^{142}$Cs case, average kinetic energies were calculated from the Fermi and Wolinska II predicted spectra (see Fig.~\ref{fig_decvCs}), resulting in values of 2.974$\pm$0.020~MeV and 2.422$\pm$0.020~MeV. The $^{92}$Rb and $^{142}$Cs average $\beta$ energy values determined from predictions using $\beta$ feedings from TAGS measurements \cite{ALGORA1,WOLINSKA} are in excellent agreement with the corresponding experimental values reported in this work, which were directly computed from the deconvolved $\beta$ spectra. Table \ref{tab_avgene} contains the reported average $\beta$ energy values for the $^{92}$Rb and $^{142}$Cs $\beta$ decays.

        \begin{table}[ht]
            \centering
            \caption{Comparison of the $^{92}$Rb and $^{142}$Cs average $\beta$ energy values obtained from the deconvoluted spectra (Decv. Spec.), extended at low energies with the best respective predictions (see Fig. \ref{fig_decvRb} and \ref{fig_decvCs}), against reference values.}
            \label{tab_avgene}
            \begin{tabular}{|c|c|}
                \hline
                \multicolumn{2}{|c|}{\textbf{$\boldsymbol{^{92}}$Rb Average $\boldsymbol\beta$ Energy}} \\
                \hline
                Decv. Spec.          & 3.52$\pm$0.03 MeV   \\
                \hline
                TAGS \cite{ALGORA1}  & 3.498$\pm$0.105 MeV \\
                \hline
                ENSDF \cite{ALGORA1} & 3.640$\pm$0.030 MeV \\
                \hline
                \multicolumn{2}{|c|}{\textbf{$\boldsymbol{^{142}}$Cs Average $\boldsymbol\beta$ Energy}} \\
                \hline
                Decv. Spec.                         & 2.41$\pm$0.01 MeV   \\
                \hline
                Wolinska II (Fig.~\ref{fig_decvCs}) & 2.422$\pm$0.020 MeV \\
                \hline
                ENSDF (Fig.~\ref{fig_decvCs}, Fermi)       & 2.974$\pm$0.020 MeV \\
                \hline
            \end{tabular}
        \end{table}

	% Conclusion
	\section{Conclusion}

        In this work, we have presented a more extended description of the first measurement of $\beta$ spectrum shapes of the $\beta$ decay of $^{92}$Rb and $^{142}$Cs using radioactive beams of high-isotopic purity. The present work complements the results presented in our Letter \cite{ALCALA_PRL}. 

        $\beta$ spectrum shape determination is a complex procedure that requires a detailed characterization of the experimental setup. The satisfactory results of the validation of the Monte Carlo model of the I233 experimental setup, presented in Fig. \ref{fig_mcval}, demonstrate the reliability of \textit{e-Shape} detectors for measuring accurate $\beta$ spectra. 

		Comparisons of the best predicted spectra with respect to the $^{92}$Rb and $^{142}$Cs deconvolved data do not indicate a clear preference between the sets of correction factors for allowed $\beta$ spectra proposed by Hayen \textit{et al.} \cite{HAYEN_Allowed} and Huber \cite{HUBER}. This is expected as the shapes of the corrections in both sets are mostly similar. The correction factors proposed by Hayen \textit{et al.} \cite{HAYEN_Allowed} can account for a broader range of decay energies, as this set includes a larger number of corrections compared to those of Huber \cite{HUBER}. The effects of the tested shape correction factors for first-forbidden transitions, calculated by Hayen \textit{et al.} \cite{HAYEN_1F}, are negligible for forbidden transitions associated with a spin change of $\Delta I^\pi = 0^-$ in the decaying nuclei. Fig. \ref{fig_decvRb_FFcor} shows that the differences in the spectra, due to the inclusion of this factor, fall inside the one-sigma error band of our measurements.

       Our study also highlights the importance of considering $\beta$ feedings free from the Pandemonium effect, as determined from TAGS experiments, to describe theoretically $\beta$ spectra of isotopes relevant for calculating reliable reactor antineutrino spectra predictions \cite{ALGORA1}.
       
       The differences in the $^{92}$Rb and $^{142}$Cs $\beta$ spectra reported in this article, compared to those of Rudstam \textit{et al.} \cite{RUDSTAM_1990}, are probably due to the improved experimental conditions and the modern tools used to analyze the data presented in this article. The analysis of the I233 experimental data benefited from the use of high-purity radioactive beams produced at the IGISOL facility, which were unavailable at the time of the measurements conducted by Tengblad \textit{et al.} \cite{TENGBLAD}. Please note that the Rudstam \textit{et al.} \cite{RUDSTAM_1990} $\beta$ spectra of interest had to be disentangled from spectra originating from isobaric $\beta$ decay chains by analyzing the $\gamma$-rays from the de-excitation of the respective daughter nuclei, which could lead to systematic contamination. Another difference with the earlier work is that, in our experiment, the \textit{e-Shape} detector responses were obtained directly from modern Monte Carlo simulations rather than being approximated using models for the responses.
       
       Future analyses of relevant $\beta$ spectra for reactor antineutrino spectra predictions will include first-forbidden transitions with $\Delta I > 0$ and the testing of their corresponding forbidden shape correction factors \cite{HAYEN_1F}.

    \begin{acknowledgments}

        This work has been supported by the Spanish Ministerio de Economía y Competitividad Grant No. FPA2017-83946-C2-1-P, by the Spanish Ministerio de Ciencia e Innovación Grants No. PID2019-104714GB-C21 and PID2022-138297NB-C21, and by the Generalitat Valenciana Prometeo Grant CIPROM/2022/9. This work has also been supported by the CNRS challenge NEEDS and the associated NACRE project, which co-founded a part of the experimental setup and A. Beloeuvre's PhD grant, the CNRS/IN2P3 \textit{e-Shape} and CNRS/IN2P3 PICS TAGS between Subatech and IFIC, and the CNRS/IN2P3 Master projects Jyväskylä and OPALE. Part of the \textit{e-Shape} detector and R. Kean's postdoc contract were funded by a grant from the Pays de Loire region. Authors also acknowledge the financial support from the Ministerio de Ciencia e Innovación with funding from the European Union NextGenerationEU and Generalitat Valenciana in the call Programa de Planes Complementarios de I+D+i (PRTR 2022) under Project DETCOM, reference ASFAE/2022/027. Partial support from the EU APRENDE project (grant agreement no. 101164596) is also acknowledged. G. A. Alcal\'a acknowledges the support of the Santiago Grisolia Program of the Comunitat Valenciana. W. Gelletly acknowledges the support of the U.K. Science and Technology Facilities Council grant ST/P005314. V. Guadilla acknowledges the support of the National Science Center, Poland, under Contract No. 2019/35/D/ST2/02081. The support of the EU Horizon 2020 research and innovation program under Grant No. 771036 (ERC CoG MAIDEN) is also acknowledged. Authors would also like to thank Neil Clark from MICRON, Massimo Volpi from CAEN, and Paul Schotanus from Scionix for fruitful discussions, as well as Leendert Hayen for sharing the beta-shape theoretical corrections with our collaboration.
        
        We especially thank Subatech members S. Bouvier and J.-S. Stutzmann for their electronic and mechanical technical support. IFIC and the Univ. of Jyv\"askyl\"a are also acknowledged for their technical aids. Enlightening discussions with Alejandro Sonzogni, Jouni Suhonen, and Marlom De Oliveira Ramalho are also acknowledged. We also thank B. C. Rasco for providing us with the published $^{92}$Rb $\beta$ decay feeding data for the comparisons.
        
    \end{acknowledgments}

    \appendix*
    \section{Spectral Data}

        Table \ref{tab_specdata} lists the numerical values of the deconvolved $^{92}$Rb and $^{142}$Cs $\beta$ spectra measured with the \textit{e-Shape} detectors during the I233 experiment, together with their corresponding errors. The values of the best model spectra that describe the experimental data are also given. The best model (or predicted) spectra were normalized to the respective experimental spectra to extend the values of the experimental data in the low-energy region, as deconvolved low-energy points are difficult to obtain due to the energy loss of low-energy $\beta$ electrons in the Si detector. The errors of the extended low-energy data points were assigned the same uncertainty as the respective first experimental data points, following the convention of Rudstam \textit{et al.} \cite{RUDSTAM_1990}.

        \begin{longtable*}{|c|c|c|c|c|c|c|c|c|c|}

            \caption{The values of the deconvolved $^{92}$Rb and $^{142}$Cs $\beta$ spectra ($S_\beta$) obtained in the I233 experiment are given in columns Exp.~$^{92}$Rb and Exp.~$^{142}$Cs, respectively. The best model $\beta$ spectra (see $\chi^2$/df values in Fig. \ref{fig_decvRb} and \ref{fig_decvCs}), generated with the Zakari~II and Wolinska~II TAGS $\beta$ feedings together with the corresponding allowed and forbidden shape correction models, are shown in columns Mod.~$^{92}$Rb and Mod.~$^{142}$Cs, respectively. Spectra are given respect to the kinetic energy of the $\beta$ electrons (E$_\beta$).}
            \label{tab_specdata} \\

            \hline
            \multicolumn{3}{|c|}{\bf Exp. $\boldsymbol{^{92}}$Rb} & \multicolumn{2}{c|}{\bf Mod. $\boldsymbol{^{92}}$Rb} & \multicolumn{3}{c|}{\bf Exp. $\boldsymbol{^{142}}$Cs} & \multicolumn{2}{c|}{\bf Mod. $\boldsymbol{^{142}}$Cs} \\
            \hline
            E$_\beta$ (MeV) & $S_\beta$ & $\Delta S_\beta$ & E$_\beta$ (MeV) & $S_\beta$ & E$_\beta$ (MeV) & $S_\beta$ & $\Delta S_\beta$ & E$_\beta$ (MeV) & $S_\beta$ \\
            \hline
            \endfirsthead

            \hline
            \multicolumn{3}{|c|}{\bf Exp. $\boldsymbol{^{92}}$Rb} & \multicolumn{2}{c|}{\bf Mod. $\boldsymbol{^{92}}$Rb} & \multicolumn{3}{c|}{\bf Exp. $\boldsymbol{^{142}}$Cs} & \multicolumn{2}{c|}{\bf Mod. $\boldsymbol{^{142}}$Cs} \\
            \hline
            E$_\beta$ (MeV) & $S_\beta$ & $\Delta S_\beta$ & E$_\beta$ (MeV) & $S_\beta$ & E$_\beta$ (MeV) & $S_\beta$ & $\Delta S_\beta$ & E$_\beta$ (MeV) & $S_\beta$ \\
            \hline
            \endhead

            \hline
            \endfoot

            \hline
            \endlastfoot

            0.05 & 0.0183    & 0.0351        & 0.05 & 0.0183        & 0.05 & 0.0724     & 0.0148         & 0.05 & 0.0724        \\
            0.15 & 0.0351    & 0.0351        & 0.15 & 0.0351        & 0.15 & 0.1144     & 0.0148         & 0.15 & 0.114         \\
            0.25 & 0.0477    & 0.0351        & 0.25 & 0.0477        & 0.25 & 0.1419     & 0.0148         & 0.25 & 0.142         \\
            0.35 & 0.0592    & 0.0351        & 0.35 & 0.0592        & 0.35 & 0.1667     & 0.0148         & 0.35 & 0.167         \\
            0.45 & 0.0734    & 0.0351        & 0.45 & 0.0701        & 0.45 & 0.1897     & 0.0148         & 0.45 & 0.190         \\
            0.55 & 0.0804    & 0.0255        & 0.55 & 0.0804        & 0.55 & 0.2143     & 0.0148         & 0.55 & 0.211         \\
            0.65 & 0.0859    & 0.0200        & 0.65 & 0.0901        & 0.65 & 0.2289     & 0.0118         & 0.65 & 0.230         \\
            0.75 & 0.0903    & 0.0181        & 0.75 & 0.0991        & 0.75 & 0.24672    & 0.00852        & 0.75 & 0.248         \\
            0.85 & 0.0995    & 0.0175        & 0.85 & 0.108         & 0.85 & 0.26445    & 0.00566        & 0.85 & 0.262         \\
            0.95 & 0.1074    & 0.0162        & 0.95 & 0.115         & 0.95 & 0.28268    & 0.00354        & 0.95 & 0.275         \\
            1.05 & 0.1170    & 0.0161        & 1.05 & 0.123         & 1.05 & 0.29591    & 0.00214        & 1.05 & 0.285         \\
            1.15 & 0.1259    & 0.0170        & 1.15 & 0.129         & 1.15 & 0.30768    & 0.00129        & 1.15 & 0.292         \\
            1.25 & 0.1330    & 0.0157        & 1.25 & 0.136         & 1.25 & 0.312798   & 0.000868       & 1.25 & 0.298         \\
            1.35 & 0.1408    & 0.0114        & 1.35 & 0.141         & 1.35 & 0.312876   & 0.000844       & 1.35 & 0.301         \\
            1.45 & 0.14894   & 0.00558       & 1.45 & 0.147         & 1.45 & 0.31118    & 0.00103        & 1.45 & 0.302         \\
            1.55 & 0.15599   & 0.00144       & 1.55 & 0.151         & 1.55 & 0.30465    & 0.00119        & 1.55 & 0.301         \\
            1.65 & 0.16069   & 0.00086       & 1.65 & 0.156         & 1.65 & 0.30038    & 0.00133        & 1.65 & 0.298         \\
            1.75 & 0.16592   & 0.00192       & 1.75 & 0.160         & 1.75 & 0.29522    & 0.00142        & 1.75 & 0.294         \\
            1.85 & 0.17036   & 0.00233       & 1.85 & 0.165         & 1.85 & 0.29048    & 0.00144        & 1.85 & 0.288         \\
            1.95 & 0.17333   & 0.00263       & 1.95 & 0.169         & 1.95 & 0.28446    & 0.00144        & 1.95 & 0.282         \\
            2.05 & 0.17509   & 0.00288       & 2.05 & 0.173         & 2.05 & 0.27493    & 0.00140        & 2.05 & 0.274         \\
            2.15 & 0.17824   & 0.00309       & 2.15 & 0.177         & 2.15 & 0.26622    & 0.00137        & 2.15 & 0.266         \\
            2.25 & 0.18246   & 0.00307       & 2.25 & 0.182         & 2.25 & 0.25826    & 0.00133        & 2.25 & 0.258         \\
            2.35 & 0.18652   & 0.00302       & 2.35 & 0.186         & 2.35 & 0.25109    & 0.00131        & 2.35 & 0.249         \\
            2.45 & 0.19143   & 0.00305       & 2.45 & 0.189         & 2.45 & 0.24411    & 0.00127        & 2.45 & 0.241         \\
            2.55 & 0.19411   & 0.00301       & 2.55 & 0.193         & 2.55 & 0.23566    & 0.00123        & 2.55 & 0.233         \\
            2.65 & 0.19663   & 0.00317       & 2.65 & 0.196         & 2.65 & 0.22617    & 0.00118        & 2.65 & 0.225         \\
            2.75 & 0.19835   & 0.00333       & 2.75 & 0.199         & 2.75 & 0.22035    & 0.00116        & 2.75 & 0.218         \\
            2.85 & 0.19992   & 0.00327       & 2.85 & 0.202         & 2.85 & 0.21282    & 0.00113        & 2.85 & 0.211         \\
            2.95 & 0.20168   & 0.00332       & 2.95 & 0.204         & 2.95 & 0.20742    & 0.00110        & 2.95 & 0.205         \\
            3.05 & 0.20445   & 0.00346       & 3.05 & 0.206         & 3.05 & 0.19874    & 0.00106        & 3.05 & 0.198         \\
            3.15 & 0.20679   & 0.00343       & 3.15 & 0.207         & 3.15 & 0.19246    & 0.00103        & 3.15 & 0.192         \\
            3.25 & 0.20811   & 0.00339       & 3.25 & 0.208         & 3.25 & 0.18727    & 0.00100        & 3.25 & 0.186         \\
            3.35 & 0.21002   & 0.00349       & 3.35 & 0.209         & 3.35 & 0.181949   & 0.000980       & 3.35 & 0.180         \\
            3.45 & 0.21109   & 0.00351       & 3.45 & 0.209         & 3.45 & 0.177889   & 0.000966       & 3.45 & 0.174         \\
            3.55 & 0.21170   & 0.00337       & 3.55 & 0.209         & 3.55 & 0.172816   & 0.000941       & 3.55 & 0.168         \\
            3.65 & 0.21046   & 0.00332       & 3.65 & 0.208         & 3.65 & 0.165597   & 0.000908       & 3.65 & 0.162         \\
            3.75 & 0.20714   & 0.00330       & 3.75 & 0.207         & 3.75 & 0.156874   & 0.000862       & 3.75 & 0.156         \\
            3.85 & 0.20420   & 0.00347       & 3.85 & 0.206         & 3.85 & 0.150430   & 0.000835       & 3.85 & 0.150         \\
            3.95 & 0.20290   & 0.00348       & 3.95 & 0.204         & 3.95 & 0.143452   & 0.000800       & 3.95 & 0.144         \\
            4.05 & 0.20058   & 0.00330       & 4.05 & 0.202         & 4.05 & 0.139064   & 0.000781       & 4.05 & 0.138         \\
            4.15 & 0.19893   & 0.00319       & 4.15 & 0.199         & 4.15 & 0.132445   & 0.000749       & 4.15 & 0.132         \\
            4.25 & 0.19361   & 0.00324       & 4.25 & 0.196         & 4.25 & 0.125228   & 0.000715       & 4.25 & 0.126         \\
            4.35 & 0.19249   & 0.00333       & 4.35 & 0.193         & 4.35 & 0.118865   & 0.000685       & 4.35 & 0.120         \\
            4.45 & 0.18867   & 0.00314       & 4.45 & 0.189         & 4.45 & 0.111605   & 0.000650       & 4.45 & 0.113         \\
            4.55 & 0.18632   & 0.00287       & 4.55 & 0.185         & 4.55 & 0.105543   & 0.000620       & 4.55 & 0.107         \\
            4.65 & 0.18068   & 0.00279       & 4.65 & 0.181         & 4.65 & 0.099943   & 0.000593       & 4.65 & 0.100         \\
            4.75 & 0.17522   & 0.00290       & 4.75 & 0.176         & 4.75 & 0.094603   & 0.000569       & 4.75 & 0.0940        \\
            4.85 & 0.17012   & 0.00296       & 4.85 & 0.171         & 4.85 & 0.087639   & 0.000537       & 4.85 & 0.0877        \\
            4.95 & 0.16490   & 0.00283       & 4.95 & 0.166         & 4.95 & 0.081688   & 0.000508       & 4.95 & 0.0814        \\
            5.05 & 0.16058   & 0.00269       & 5.05 & 0.160         & 5.05 & 0.074596   & 0.000473       & 5.05 & 0.0752        \\
            5.15 & 0.15455   & 0.00252       & 5.15 & 0.155         & 5.15 & 0.068189   & 0.000441       & 5.15 & 0.0691        \\
            5.25 & 0.14912   & 0.00246       & 5.25 & 0.149         & 5.25 & 0.061418   & 0.000407       & 5.25 & 0.0631        \\
            5.35 & 0.14331   & 0.00235       & 5.35 & 0.142         & 5.35 & 0.057122   & 0.000387       & 5.35 & 0.0574        \\
            5.45 & 0.13595   & 0.00227       & 5.45 & 0.136         & 5.45 & 0.050379   & 0.000351       & 5.45 & 0.0518        \\
            5.55 & 0.13185   & 0.00232       & 5.55 & 0.130         & 5.55 & 0.046346   & 0.000332       & 5.55 & 0.0466        \\
            5.65 & 0.12616   & 0.00224       & 5.65 & 0.123         & 5.65 & 0.043877   & 0.000322       & 5.65 & 0.0416        \\
            5.75 & 0.12225   & 0.00207       & 5.75 & 0.116         & 5.75 & 0.037112   & 0.000284       & 5.75 & 0.0370        \\
            5.85 & 0.11646   & 0.00185       & 5.85 & 0.110         & 5.85 & 0.034388   & 0.000272       & 5.85 & 0.0328        \\
            5.95 & 0.10910   & 0.00170       & 5.95 & 0.103         & 5.95 & 0.030218   & 0.000248       & 5.95 & 0.0290        \\
            6.05 & 0.10176   & 0.00168       & 6.05 & 0.0960        & 6.05 & 0.027612   & 0.000237       & 6.05 & 0.0255        \\
            6.15 & 0.09479   & 0.00164       & 6.15 & 0.0891        & 6.15 & 0.022425   & 0.000206       & 6.15 & 0.0221        \\
            6.25 & 0.08729   & 0.00148       & 6.25 & 0.0823        & 6.25 & 0.019567   & 0.000191       & 6.25 & 0.0189        \\
            6.35 & 0.08014   & 0.00132       & 6.35 & 0.0755        & 6.35 & 0.016324   & 0.000170       & 6.35 & 0.0159        \\
            6.45 & 0.07245   & 0.00118       & 6.45 & 0.0688        & 6.45 & 0.012828   & 0.000147       & 6.45 & 0.0130        \\
            6.55 & 0.06580   & 0.00106       & 6.55 & 0.0622        & 6.55 & 0.010223   & 0.000130       & 6.55 & 0.0104        \\
            6.65 & 0.058016  & 0.000935      & 6.65 & 0.0557        & 6.65 & 0.008569   & 0.000122       & 6.65 & 0.00799       \\
            6.75 & 0.050214  & 0.000847      & 6.75 & 0.0494        & 6.75 & 0.0051582  & 0.0000889      & 6.75 & 0.00587       \\
            6.85 & 0.043783  & 0.000803      & 6.85 & 0.0434        & 6.85 & 0.0034112  & 0.0000715      & 6.85 & 0.00403       \\
            6.95 & 0.038834  & 0.000703      & 6.95 & 0.0375        & 6.95 & 0.0020470  & 0.0000550      & 6.95 & 0.00251       \\
            7.05 & 0.032613  & 0.000522      & 7.05 & 0.0320        & 7.05 & 0.0012771  & 0.0000467      & 7.05 & 0.00133       \\
            7.15 & 0.026463  & 0.000414      & 7.15 & 0.0267        & 7.15 & 0.0001912  & 0.0000126      & 7.15 & 0.000506      \\
            7.25 & 0.021781  & 0.000387      & 7.25 & 0.0218        & 7.25 & 0.00003210 & 0.00000424     & 7.25 & 0.0000688     \\
            7.35 & 0.016746  & 0.000318      & 7.35 & 0.0174        & -    & -           & -             & -    & -             \\
            7.45 & 0.013510  & 0.000237      & 7.45 & 0.0133        & -    & -           & -             & -    & -             \\
            7.55 & 0.009359  & 0.000137      & 7.55 & 0.00968       & -    & -           & -             & -    & -             \\
            7.65 & 0.0060373 & 0.0000781     & 7.65 & 0.00658       & -    & -           & -             & -    & -             \\
            7.75 & 0.0039476 & 0.0000508     & 7.75 & 0.00403       & -    & -           & -             & -    & -             \\
            7.85 & 0.0016027 & 0.0000275     & 7.85 & 0.00207       & -    & -           & -             & -    & -             \\
            7.95 & 0.0006041 & 0.0000180     & 7.95 & 0.000736      & -    & -           & -             & -    & -             \\
            8.05 & 0.0002043 & 0.0000119     & 8.05 & 0.0000713     & -    & -           & -             & -    & -             \\
            
        \end{longtable*}

    \bibliography{bib_I233_PRC}{}
    \bibliographystyle{apsrev4-2}

\end{document}